\documentclass[sigconf]{aamas} 
\usepackage{tabularx}

\usepackage{balance} 

\setcopyright{ifaamas}
\acmConference[AAMAS '24]{Proc.\@ of the 23rd International Conference
on Autonomous Agents and Multiagent Systems (AAMAS 2024)}{May 6 -- 10, 2024}
{Auckland, New Zealand}{N.~Alechina, V.~Dignum, M.~Dastani, J.S.~Sichman (eds.)}
\copyrightyear{2024}
\acmYear{2024}
\acmDOI{}
\acmPrice{}
\acmISBN{}

\acmSubmissionID{537}

\title[AAMAS-2024 Formatting Instructions]{First 100 days of pandemic; an interplay of pharmaceutical, behavioural and digital interventions - A study using agent based modeling}

\author{Gauri Gupta}
\affiliation{
  \institution{Massachusetts Institute of Technology}
  \city{Cambridge, Massachusetts}
  \country{United States}}
\email{gaurii@mit.edu}

\author{Ritvik Kapila}
\affiliation{
  \institution{University of California San Diego}
  \city{San Diego, California}
  \country{United States}}
\email{rkapila@ucsd.edu}

\author{Ayush Chopra}
\affiliation{
  \institution{Massachusetts Institute of Technology}
  \city{Cambridge, Massachusetts}
  \country{United States}}
\email{ayushc@mit.edu}

\author{Ramesh Raskar}
\affiliation{
  \institution{Massachusetts Institute of Technology}
  \city{Cambridge, Massachusetts}
  \country{United States}}
\email{raskar@mit.edu}

\begin{abstract}

Pandemics, notably the recent COVID-19 outbreak, have impacted both public health and the global economy. A profound understanding of disease progression and efficient response strategies is thus needed to prepare for potential future outbreaks. In this paper, we emphasize the potential of Agent-Based Models (ABM) in capturing complex infection dynamics and understanding the impact of interventions. We simulate realistic pharmaceutical, behavioral, and digital interventions that mirror challenges in real-world policy adoption and suggest a holistic combination of these interventions for pandemic response. Using these simulations, we study the trends of emergent behavior on a large-scale population based on real-world socio-demographic and geo-census data from Kings County in Washington. Our analysis reveals the pivotal role of the initial 100 days in dictating a pandemic's course, emphasizing the importance of quick decision-making and efficient policy development. Further, we highlight that investing in behavioral and digital interventions can reduce the burden on pharmaceutical interventions by reducing the total number of infections and hospitalizations, and by delaying the pandemic's peak. We also infer that allocating the same amount of dollars towards extensive testing with contact tracing and self-quarantine offers greater cost efficiency compared to spending the entire budget on vaccinations.

\end{abstract}

\keywords{Agent-based Models; Computational Epidemiology; Sensitivity Analysis; Contact Tracing Intervention; COVID-19 Pandemic}

\newcommand{\BibTeX}{\rm B\kern-.05em{\sc i\kern-.025em b}\kern-.08em\TeX}

\begin{document}


\pagestyle{fancy}
\fancyhead{}

\maketitle
 
\section{Introduction}
The recent outbreaks of COVID-19 have left an indelible mark on society at a global scale, highlighting the vulnerability of public health \cite{miyah2022covid, lewis2022covid}. Thus, deepening our insights into how pandemics evolve is imperative, ensuring that our actions are prompt, effective, and grounded in evidence \cite{amzat2020coronavirus, tam2020use}. Given the unprecedented nature of these pandemics, it is inherently challenging to simulate their dynamics. Agent-based modeling has emerged as a pivotal tool for replicating the complex dynamics inherent in the pandemic evolution \cite{romero2021public, abueg2021modeling, aleta2020modelling, kerr2021covasim}. Agent-based models (ABMs) are distinct in their ability to provide a granular view of disease propagation by analyzing both micro-level interactions and the broader emergent phenomena, making them particularly suited for delineating the effects of potential interventions. 

In the past, governments globally adopted varied strategies to counteract the spread of infections, particularly during COVID-19 \cite{haug2020ranking, brauner2021inferring}. Some of these were effective, while some were not \cite{flaxman2020estimating, banholzer2021estimating}. Interventions such as delayed travel bans proved insufficient, allowing for rapid global infection spread \cite{tellis2020price}, while prolonged severe lockdowns crippled global economies \cite{moshe2022lockdowns}. Additionally, the deployment of digital initiatives for contact tracing \cite{cheng2020contact, raymenants2022empirical, willem2021impact, barrat2021effect} had a limited impact due to low adoption and delays in user quarantine post-exposure \cite{davis2021contact}. As notified users awaited test results, potential carriers inadvertently continued activities, making this approach largely ineffective in curbing transmission \cite{li2020covid, chowdhury2020covid}. Pharmaceutical interventions, once viewed as the primary defense against the pandemic, encountered their own set of challenges \cite{xiaopharmadiff}. The apprehension over the longevity of vaccine-induced immunity and potential side effects further hampered the pace of vaccination drives \cite{subbarao2021success, odone2020vaccine}. Moreover, by the time effective vaccines were produced, many nations had already peaked in infections \cite{pandey2021challenges}. Thus, reflecting on these previous strategies to understand what worked and what did not is crucial for developing efficient future pandemic responses.


However, decision-making in such scenarios is challenging due to the multifarious intricacies of complex societies characterized by heterogeneous populations, diverse behavioral patterns, and differential access to resources \cite{alsalem2022multi, angeli2020sensemaking, chang2021mobility}. The interplay of various interventions, their mutual impacts, and the factors influencing their effectiveness adds further layers of complexity. In this paper, we address these challenges of modeling real-world simulations in complex societies by considering varied populations with interaction networks spanning across household, occupational, and random graphs. We consider behavioral patterns through app adoption rates, self-quarantine, and compliance probabilities. Further, we model differential access to resources stratified by age or policy choices, for instance, prioritizing higher age group individuals for vaccination and age-based app adoption. 

We model the progression of a pandemic over its initial 6 months (180 days) using real-world socio-demographic (census) data from Kings County, Washington, evaluated at scale of 100,000 agents. For our analysis, we simulate three main types of interventions: pharmaceutical, behavioral, and digital, and highlight the effectiveness and potential pitfalls of each approach in controlling future pandemics. We not only assess these individual interventions but also integrate a collective interplay of these interventions, suggesting they are complementary to each other, not alternate. Pharmaceutical interventions include vaccination drives and testing to detect cases. Behavioral interventions include self-quarantine upon testing positive and individual responsiveness and adherence to recommended actions. The lockdowns many countries implemented can be viewed as a prolonged strict self-quarantine. Digital interventions explore tools such as contact tracing apps designed to monitor and curtail spread through tracking interactions. Our extensive analysis suggests relying solely on rapid vaccine development for outbreak control isn't viable. Enhanced preparedness demands an integration of pharmaceutical approaches with contact tracing and behavioral strategies, ensuring a holistic, prompt response. 

The following are the major contributions of our paper:
(1) We introduce a general pipeline using ABMs that simulates a real-world synergy of interventions at scale, encompassing pharmaceutical, behavioral, and digital strategies. This framework offers extensive detail to capture the complexities observed in the real-world adoption of these interventions. (2) Our user-friendly and flexible framework is designed with a customizable configuration file enabling non-technical people like epidemiologists and policy-makers to study the effect of intricate interventions on pandemics. (3) We provide a comprehensive cost analysis of pandemic containment under each intervention strategy. (4) We perform extensive experiments on real-world data from Kings County, Washington for COVID-19. Our findings deepen the understanding of pandemic trends and offer valuable policy recommendations for effective pandemic response. (5) Some of our interesting insights are: (a) The first 100 days of the pandemic are a pivotal threshold in determining the course of a pandemic's trajectory. (b) Pairing delayed vaccination with digital and behavioral interventions proves more impactful than solely pushing for early vaccination, as it not only reduces overall infections and hospitalizations but also delays their peak. (c) With a fixed \$0.5M budget, investing in testing with self-quarantine and digital contact tracing is more effective than funding early vaccinations alone.

\section{Related Work}
Agent-based models (ABMs) are discrete simulators that allow entities (agents) with designated characteristics to interact within a given computational environment, replicating complex systems \cite{bonabeau2002agent, norton2019multiscale, reiker2021machine, zheng2022ai, dimitrov2010mathematical, hethcote2000mathematics}. Recently, ABMs have been widely employed in epidemiology to understand disease progression and the efficacy of interventions by providing relevant information to investigate and predict the behavior of the pandemic \cite{marathe2013computational, romero2021public, abueg2021modeling, aleta2020modelling, kerr2021covasim}. Several studies have utilized ABMs to evaluate the effectiveness of different interventions, such as social distancing, quarantine, lockdown, and vaccination \cite{hinch2021openabm, chopra2021deepabm, 
romero2021public}. ABMs have also been used in prior works for addressing policy-related queries like evaluating the importance of test turnaround time versus its sensitivity \cite{larremore2021test}, and the benefits of postponing the vaccine's second dose to focus on the distribution of the first dose \cite{romero2021public}.

However, the utility of ABMs for practical decision-making depends upon several factors. These include their accuracy in replicating the population behavior \cite{pellis2015eight, groff2019state}. Furthermore, ABMs are conventionally slow. A single forward simulation over a large ABM can take several days \cite{aylett2021june, bisset2009epifast}. ABM simulations are difficult to scale to large populations \cite{bisset2009epifast}, and are tough to calibrate with real-world data \cite{pellis2015eight}. Moreover, they only simulate the effect of one intervention at a time, where real-world deployment of intervention strategies are intricately linked to each other and should be studied with a combined effect of each of these \cite{kerr2021covasim, chopra2021deepabm, hinch2021openabm}. 

Only after overcoming these challenges in ABMs can their insights truly guide strategic pandemic interventions. Our model provides a comprehensive system that simulates interventions with real-world challenges of deployment or adoption, representing them through quantifiable parameters. We adopt a vectorized approach \cite{chopra2021deepabm} which enables a fast, parallelized simulation, allowing analysis of emergent behavior on a large-scale population for millions of agents in a few seconds. We not only assess the individual interventions but also integrate a holistic interplay of these interventions.


\section{Method}
\begin{figure*}
    \begin{center}
    \includegraphics[width=\linewidth]{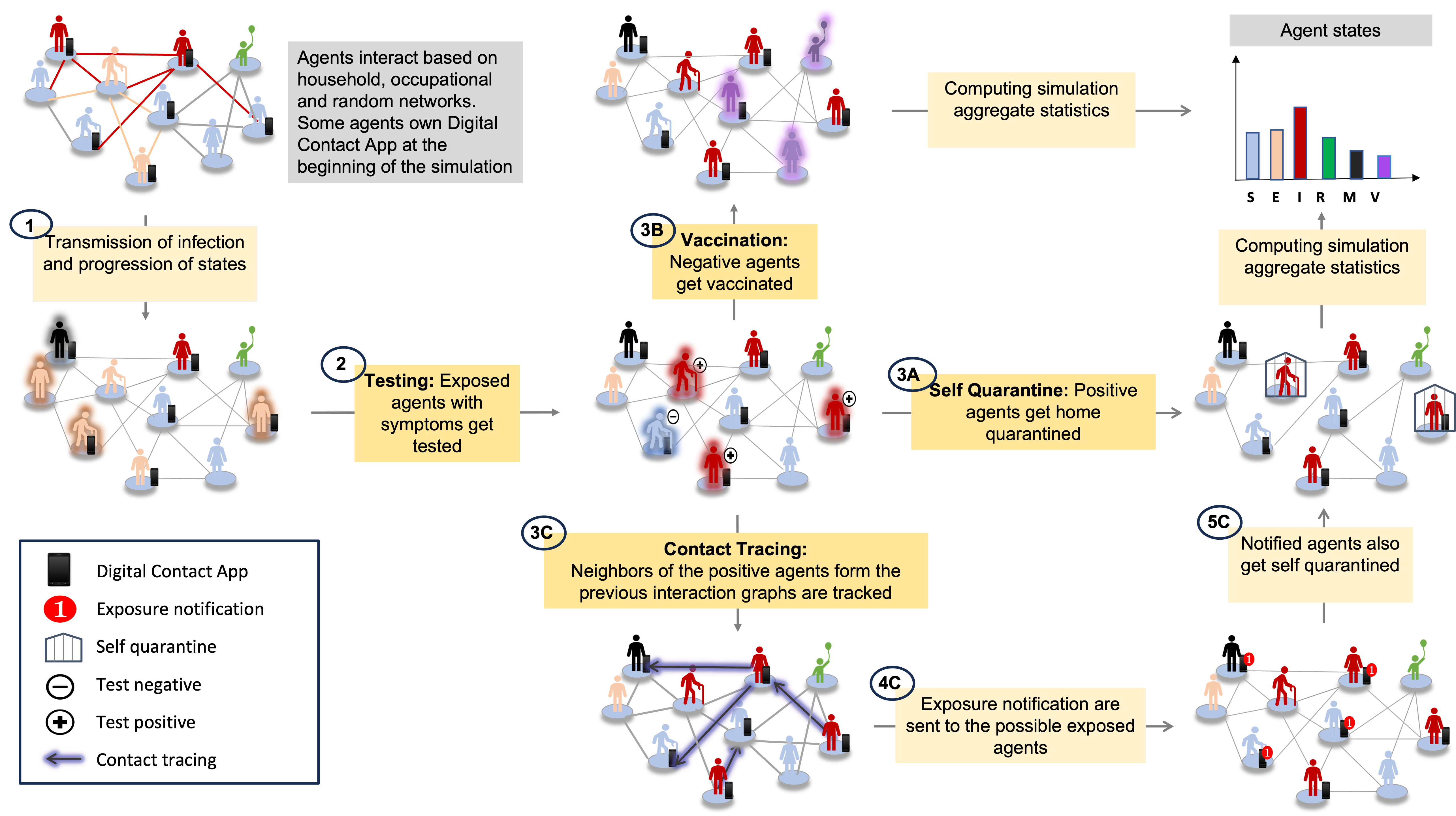}
    \caption{Implementation of different interventions - Testing, Self-quarantine, Vaccination, and Contact Tracing. (1) Infection spreads through the interaction of infected with susceptible agents, and the states of the agents are then updated based on disease progression. (2) Upon experiencing symptoms, exposed agents get themselves tested (3a) If tested positive, agents undergo self-quarantine with compliance. A quarantined agent then engages in no further interactions until the quarantine period ends. The interaction graph of quarantine agents is thus an isolated point (3b) Agents that have not tested positive or are not quarantined get vaccinated. Vaccination reduces the susceptibility of an agent to infection risk (3c) In case of contact tracing: interactions of the positively tested agents (that own app in case of DCT) from the previous interaction graphs of past days are tracked; (4c) exposure notifications are sent to the possibly exposed tracked agents (that own the app in case of DCT); (5c) notified agents then opt for self-quarantine. (Last) After simulating for N days, the aggregate statistics of the agent states are computed. Agent states here are: susceptible (S), exposed (E), infected (I), recovered (R), mortal (M), and vaccinated (V)}
    \label{fig:pipeline}
    \end{center}
\end{figure*}

Figure \ref{fig:pipeline} shows the pipeline for different interventions along with the progression of disease stages. We build on existing open-source agent-based modeling frameworks \cite{chopra2021deepabm, hinch2021openabm}, optimizing large-scale simulations through matrix computations, leading to enhanced computational efficiency. Our model addresses the potential variances in the adoption of interventions by employing a stochastic approach, sampling from Gaussian distributions based on a certain compliance level to predict outcomes. We model these interventions with an unprecedented level of detail with aspects that have not been explored in such granular detail in prior research. This comprehensive approach not only facilitates the examination of the individual impact of each parameter, but also provides insights into the synergistic effects of these interventions on the pandemic response.

In our model, individual agents (and their states) are modeled as tensors. Agents navigate through eleven potential disease stages: susceptible, asymptomatic, presymptomatic (mild or severe), symptomatic (mild or severe), hospitalized, in intensive care, recovered, immunized, or deceased. Infections can propagate during any interaction between susceptible and infected agents. These interactions span three networks: household, occupation, and random encounters, which are represented as sparse adjacency matrices. Each such interaction is a stochastic process with a certain risk of disease transmission. The foundational assumptions about disease progression, transmission dynamics, and network interactions align with prior agent-based models \cite{chopra2021deepabm, romero2021public}. We will now delve into an in-depth examination of each intervention, detailing various parameters and compliance factors reflective of real-world scenarios.

\subsection{Testing}
Agents who are exposed to infection and develop symptoms undergo testing. Every diagnostic test is defined by three primary parameters: specificity, turnaround time, and the duration of test validity. The turnaround time accounts for any inherent delays in receiving test results, presented as a dictionary detailing possible result dates and their associated probabilities. The test validity indicates the duration for which the test results are considered relevant. After this period, agents are expected to be retested. Factoring in real-world delays related to the deployment of testing kits, tests can be deployed in the model after some start date, marking the start of distribution of that particular testing method to the public. Table \ref{tab:testing_parameters} shows the list of different parameters supporting the testing mechanism.

In our simulations, we employ two types of tests: (i) RT-PCR test, with a specificity of 0.99 and a turnaround time of 1 to 3 steps (1 to 3 days) uniformly distributed \cite{chopra2021deepabm, kostoulas2021diagnostic}, and (ii) rapid point-of-care test, which offers slightly reduced specificity of 0.85 with a turnaround time of 0 steps (same day). To cater to varying diagnostic requirements, these parameters can be adjusted, offering flexibility in modeling different test types. By default, our model uses the more reliable RT-PCR test for simulations unless specified otherwise.
\begin{table}[t]
    \centering
    \caption{Description of Testing parameters}
    \label{tab:testing_parameters}
    \begin{tabular}{p{3cm}p{4.8cm}} 
        \toprule
        \textbf{Parameter} & \textbf{Explanation} \\ \midrule
        test\_start\_date & Date on which testing begins \\
        test\_true\_positive & Prob. of a true positive result \\
        test\_false\_positive & Prob. of a false positive result \\
        test\_results\_dates & Potential dates of receiving test results \\
        test\_results\_dates\_probs & Dictionary of probabilities associated with each test result date \\
        test\_validity\_days & Duration for test results validity \\
        test\_cost & Average cost of production of a test \\ \bottomrule
    \end{tabular}
\end{table}

\subsection{Self-quarantine(SQ)}
Upon testing positive or receiving exposure notification, an agent undergoes a 14-day self-quarantine adhering to compliance. However, an agent might not consistently adhere to the complete quarantine. To simulate such imperfections, a daily dropout probability of 1\% is incorporated to account for potential non-compliance. During the quarantine period, the agent's interaction network effectively becomes isolated, leading to no interactions with other agents. After successfully completing the quarantine period, the agent's capacity to transmit the infection is nullified, effectively resetting their infectiousness to zero. Table \ref{tab:sq} provides a detailed overview of different parameters for modeling self-quarantine.
\begin{table}[t]
    \centering
    \caption{Description of Self-quarantine parameters}
    \label{tab:sq}
    \begin{tabular}{p{2.3cm}p{5.5cm}}
    \toprule
        \textbf{Parameter} & \textbf{Explanation} \\ \midrule
        quar\_enter\_prob & Prob. with which an agent enters self-quarantine after testing positive \\
        quar\_break\_prob & Daily quarantine dropout probability due to non-compliance \\
        quar\_days & Number of self-quarantine days \\ \bottomrule
    \end{tabular}
\end{table}    

\subsection{Vaccination(VACC)}
We simulate a two-dose vaccination regimen with an extensive level of granularity. A dose of the vaccine provides a certain probability of becoming immune to infections, depending on whether it is a first or second dose. Vaccines are administered in an age-prioritized fashion, with the oldest individuals receiving their vaccines first, and first-dose candidates given precedence over the second dose. We simulate a probabilistic immunity conferred post-vaccination, where immunity is not immediate but materializes after a stipulated delay post-inoculation. Table \ref{tab:vaccine_parameters} illustrates the parameters driving our vaccination models, such as the vaccine's start date, daily production rate, shelf life, and efficacy percentages for both doses. Moreover, it also highlights potential dropouts - those who, after receiving the first dose might choose not to return for the second, capturing a real-world nuance in the vaccination process. 

Based on Pfizer and Moderna's trial data \cite{self2021comparative}, we assume a 90\% efficacy for the first dose and a 95\% efficacy for the second dose administered 21 days later. All simulations, unless indicated otherwise, presume vaccination commencement at t=10 with a daily vaccination rate of 0.3\% on a population of 100K based on U.S. vaccination rates and patterns observed internationally \cite{chopra2021deepabm}.
\begin{table}[t]
    \centering
    \caption{Description of Vaccine-related parameters}
    \label{tab:vaccine_parameters}
    \begin{tabular}{p{2.4cm}p{5.4cm}}
    \toprule
        \textbf{Parameter} & \textbf{Explanation} \\ \midrule
        vacc\_start\_date & Date on which vaccine drive begins \\
        vacc\_daily\_prod & No. of vaccine doses produced daily \\
        vacc\_shelf\_life & Duration before a vaccine dose expires \\
        vacc\_dose\_delay & Days after which the vaccine dose starts showing effect \\
        vacc\_dose1\_priority & Indicator if the first dose is prioritized over second in distribution \\
        vacc\_dose1\_eff & Efficacy of the first vaccine dose \\
        vacc\_dose2\_gap & Duration between the first and second doses of the vaccine \\
        vacc\_dose2\_eff & Efficacy of the second vaccine dose \\
        vacc\_dose2\_drop & Probability of an individual not returning for the second dose \\ 
        vacc\_price & Cost for development of a single vaccine \\ \bottomrule
    \end{tabular}
\end{table}

\subsection{Contact Tracing (CT)}
We adopt a hybrid contact tracing approach where first exposure notifications are dispatched to contacts of infected app users followed by manual follow-up for non-compliant users and agents not owning the app. Below, we delve into the specifics of digital and manual tracing methods.

\textbf{Digital Contact Tracing (DCT): }At the start of the simulations, agents own a Digital Contact App (DCA) with a fixed adoption rate based on age-stratified data. This app records interactions of an agent across all three networks: household, occupation, and random, within a 7-day window. Note that interactions are logged only if both agents have the app. When an agent with an app tests positive, they can opt to notify exposed contacts via the DCA. Recipients then undergo self-quarantine based on their compliance probability. In our experiments, we simulate DCT assuming an average 40\% app adoption rate and 80\% compliance rate for self-quarantine.

\begin{figure}
    \begin{center}
    \includegraphics[width=0.5\textwidth, height=0.4\textwidth]{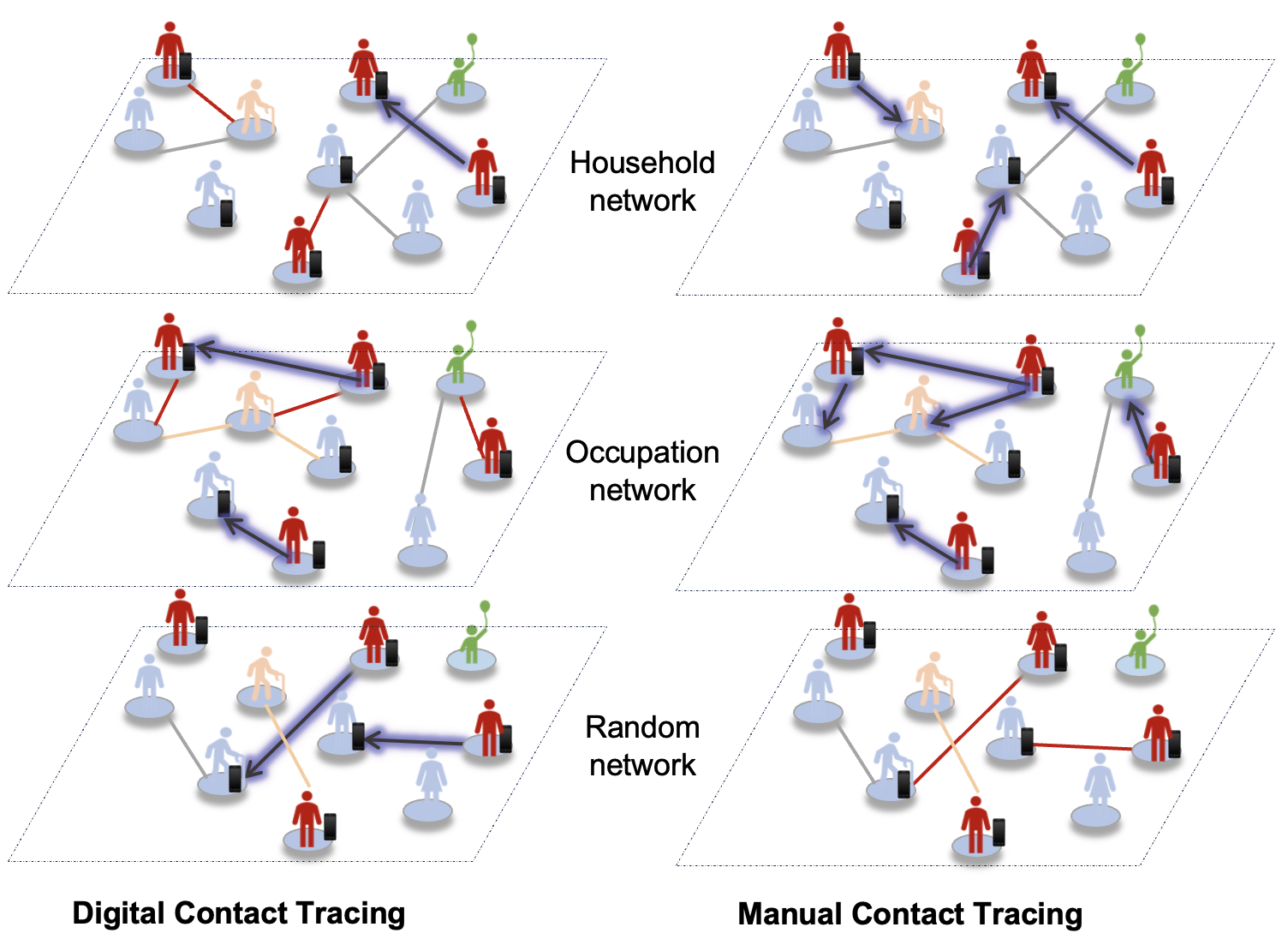}
    \caption{Comparison of Digital vs. Manual Contact Tracing: Digital tracing requires app ownership for both interacting agents but can effectively track unknown or random interactions, while manual tracing captures household and occupational contacts but may miss random interactions}
    \label{fig:mct_vs_dct}
    \end{center}
\end{figure}

\textbf{Manual Contact Tracing (MCT): }Manual tracing is similar to its digital counterpart, with a few key differences as illustrated in Figure \ref{fig:mct_vs_dct}. Unlike DCT, MCT doesn't require smartphone ownership and is unlikely to remember random or casual encounters (like those in public transport or stores). Only contacts within the household and occupational networks are traced through MCT. Manual tracers interview an infected agent to identify and track the potential contacts over the past N (=7) days. However, only a portion of the true interactions are identified based on the likelihood of recalling them (70\%). From these, a subset responds based on a set probability. Successfully contacted agents then self-quarantine with a compliance probability of 90\%.

In a targeted two-step process of contact tracing, MCT and DCT leverage coupled capabilities of human intervention with digital tools. Performing manual contact tracing of targeted potential infected agents who either did not own the app or ignored the digital notifications can significantly improve the scale/outreach of the tracing efforts to contain the infection spread. Table \ref{tab:ct_parameters} details the parameters used to model contact tracing. 
\begin{table*}[t]
    \caption{Description of Contact Tracing parameters}
    \label{tab:ct_parameters}
    \begin{tabular}{ll}
    \toprule
        \textbf{Parameter} & \textbf{Explanation} \\ \midrule
        app\_adoption\_rate & Prob. of agents owning the app at the start of the simulation in DCT\\
        max\_contact\_days & Number of days for which history of previous interactions are traced (unique for DCT and MCT)\\
        test\_inform\_prob & Prob. to notify the contacts via DCA/MCT after testing positive (unique for DCT and MCT)\\
        mct\_recall\_prob & Probab. that an individual recalls their contacts accurately during MCT \\
        mct\_reachable\_prob & Probability that an individual is reachable for manual contact tracing \\
        sq\_comply\_prob & Compliance prob. for quarantine upon succesful contact tracing (unique for DCT and MCT) \\ \bottomrule
    \end{tabular} 
\end{table*}
\section{Results}
\label{results}
We study the impact of different interventions discussed above on disease progression and pandemic evolution in a population with 100,000 agents over a period of 180-time steps. In particular, we simulate Self-Quarantine (SQ), Vaccination (VACC), and Contact Tracing (CT) interventions and evaluate their outcomes. We ran experiments using real-world socio-demographic and geo-census data from Kings County in Washington state.  All the results correspond to the mean and standard deviation aggregated over 10 independent runs of the simulation. 


We present our results in five sections. Section \ref{subsec:individual-impact} examines the individual effect of different interventions on the evolution of the pandemic outcomes. Section \ref{subsec:age-analysis} focuses on the age-stratified analysis for these individual interventions. Section \ref{subsec:cost-analysis} delves into the overall cost analysis of individual interventions, highlighting their financial implications. Section \ref{subsec:coupled-analysis} provides insights into the interplay of pharmaceutical, behavioral, and digital interventions, allowing us to study their cumulative effect on the pandemic's trajectory. Section \ref{subsec:geographical-spread} shows the geographical progression of infections in Kings County, WA, where we simulate a combination of all the interventions together and compare the spread with the unmitigated no-intervention (NI) case.

\subsection{Analysis of individual impact of different interventions} \label{subsec:individual-impact}

\begin{figure*}
    \begin{center}        \includegraphics[width=0.32\textwidth,height=0.24\textwidth]{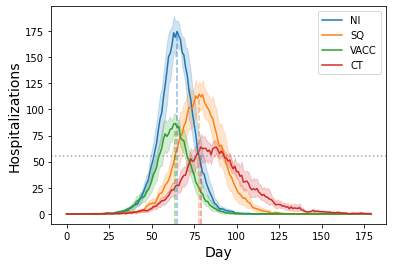}       \includegraphics[width=0.32\textwidth,height=0.24\textwidth]{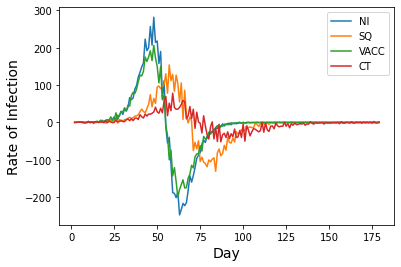}    \includegraphics[width=0.32\textwidth,height=0.24\textwidth]{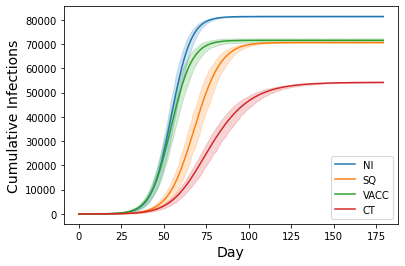}
    \caption{Comparative analysis of the individual impact of different interventions on pandemic progression;
    No Interventions (NI), Self-Quarantine (SQ), Vaccination (VACC), and Contact Tracing (CT).  (a) Peak hospitalizations showcase the strain on healthcare under each scenario, with notable stress in the NI and SQ cases. The dotted line represents the hospital bed availability for Kings County, Washington (b) Daily new infection rates highlight the efficacy of interventions, with CT significantly lowering the infection rate. (c) Cumulative infections over time reveal the pervasive nature of the pandemic in the absence of effective measures and a substantial reduction in total infections under VACC, SQ, and CT.}
    \label{fig:comparison_inter}
    \end{center}
\end{figure*}

To provide a baseline for the impact of the pandemic, we first investigate an unmitigated scenario in which there are no interventions (NI). We compare this baseline unmitigated scenario with interventions such as self-quarantine (SQ), vaccination (VACC), and contact tracing (CT). Figure \ref{fig:comparison_inter} details the comparative analysis of the individual effect of each of these interventions in terms of the number of severely affected individuals who require hospitalization, rate of infection, and cumulative infections. 

The uncontrolled pandemic (NI) peaks at t=65, with the number of hospitalizations of 175, far exceeding the available capacity of 55 \cite{covidhospcapacity, covidbedsperc} by 218\% as depicted in Figure \ref{fig:comparison_inter}(a). This indicates the immense strain on the healthcare system in the no-intervention case, pushing it to the brink of collapse. Figure \ref{fig:comparison_inter}(b) highlights a peak daily infection rate of 281 per 100,000 agents in NI, with a staggering 81\% of the population infected by the end as visualized in \ref{fig:comparison_inter}(c).

In the SQ scenario, while the maximum rate of infections dropped by 45\% compared to the NI case, hospitalizations still peaked at 115, overshooting the capacity by 109\%. This suggests that self-quarantine alone without the support of additional containment strategies is not a viable option. The VACC strategy resulted in hospitalizations still peaking at 57\% above the estimated capacity, with daily infection rates nearing those in the NI scenario. For both SQ and VACC strategies, around 70\% of the population was infected by the end of the pandemic. Interestingly, despite VACC's high infection rate, we observe fewer individuals needed hospitalization due to enhanced immunity gained by the agents through vaccination.

In the case of contact tracing (CT), a huge reduction in hospitalizations from the NI case is observed, bringing the peak very close (within 16\%) to the available capacity. Additionally, CT delayed the peak by 14 days, giving the healthcare system more time to be prepared with the necessary resources. The maximum rate of infections dropped by a massive 72\%, with only 54\% of the total population infected by the end of the pandemic. Therefore, our experiments indicate that CT with testing is the most effective standalone intervention for pandemic containment.

Notably, irrespective of the specific intervention deployed, the peak consistently occurs within the first 100 days for each individual strategy.





\subsection{Age stratification of infections for different interventions} \label{subsec:age-analysis}

\begin{figure*}
    \begin{center}
    \includegraphics[width=0.99\textwidth, height=0.3\textwidth]{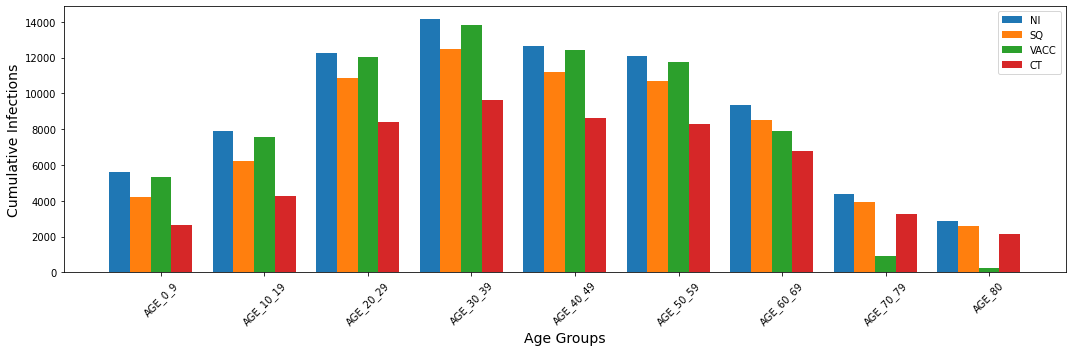}
    \caption{Age-stratified cumulative infections in Kings County, Washington, illustrating the impact of contact tracing (CT), self-quarantine (SQ), and vaccination (VACC) intervention scenarios on different age groups.}
    \label{fig:age-stratification}
    \end{center}
\end{figure*}
Figure \ref{fig:age-stratification} depicts the age-stratified cumulative infections in Kings County, Washington. While implementing CT, we simulate the app distribution with an average overall app adoption rate of 40\% in an age-stratified manner, where the age groups 20-59 have a higher probability of owning the app. Consequently, we observed a large drop of 26\% in cumulative infections for agents in these age groups of 20-59. However, a significant drop (approximately 40\%) in cumulative infections in the age group 0-19, was also observed even with relatively low app adoption rates. This is due to effective manual contact tracing implementation within households.

For the VACC intervention case, we prioritize agents in higher age groups for vaccinations. Hence, we observe a reduction of 82\% and 92\% infections for the age groups 70-79 and 80-89, respectively, as compared to the NI case. These drops are substantially high compared to the average drop in infections over all age groups of 14\%.



\subsection{Cost analysis of individual interventions} \label{subsec:cost-analysis}
\begin{figure}
    \begin{center}
    \includegraphics[width=0.4\textwidth, height=0.24\textwidth]{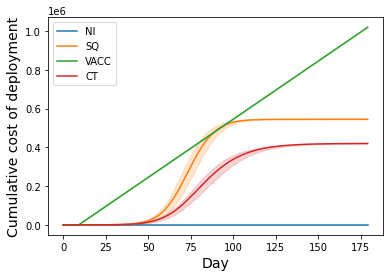}
    \caption{Comparison of costs associated with different intervention strategies. The figure shows contact tracing (CT) is the most cost-effective over both self-quarantine (SQ) and vaccination (VACC); excluding \$0 cost for no-intervention (NI)}
    \label{fig:cost_inter}
    \end{center}
\end{figure}
In this section, we evaluate the economic implications of various interventions in controlling the pandemic. The cost of the no-intervention (NI) case is \$0. For the self-quarantine (SQ) and contact tracing (CT) interventions, we account for the cost of tests taken by agents experiencing COVID-19 symptoms. The average cost per test for each case is assumed to be \$5 \cite{du2021comparative}. We assume this is the average cost per testing kit incurred by the government. Similarly, for vaccination (VACC), each dose is priced at an average of \$20 \cite{kates2023much}. 

Figure \ref{fig:cost_inter} shows the respective costs of each intervention strategy, with their individual impacts elaborated in Section \ref{subsec:individual-impact}. \ref{subsec:individual-impact}. Computing the total expenditure, VACC stands at \$1.02M, SQ at \$0.54M, and CT at a minimum of \$0.42M Beyond the reduction in infections and hospitalizations explored in Section \ref{subsec:individual-impact}, it's evident that CT surpasses by being 23\% and 59\% more cost-effective than SQ and VACC, respectively.

\begin{figure}
    \begin{center}
    \includegraphics[width=0.4\textwidth, height=0.24\textwidth]{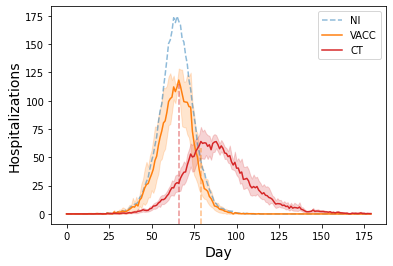}
    \caption{Comparative analysis of hospitalizations under a fixed budget of \$0.42M for contact tracing (CT) versus vaccination (VACC). The figure shows CT leads to a significant reduction in hospitalizations compared to VACC
    along with a pronounced delay in the peak, underlining the superior cost-effectiveness and strategic value of contact tracing in the pandemic's early stages.}
    \label{fig:cost_budget}
    \end{center}
\end{figure}
Further, for a fixed budget of \$0.42M, our analysis shows that deploying contact tracing with self-quarantine (CT) outperforms an exclusive focus on vaccination. In this context, the CT strategy remains consistent with the previous simulations, with only the daily vaccination production adjusted to fit the budget. As Figure \ref{fig:cost_budget} demonstrates, allocating the budget to testing with contact tracing (CT) results in a significant 63\% decline in peak hospitalizations against the no-intervention (NI) baseline. Conversely, directing the entire same budget towards vaccinations alone (VACC) yields just a 32\% reduction in peak hospitalizations compared to NI. Notably, while the VACC and NI peaks coincide, the CT approach introduces a 13-day delay in the surge of hospitalizations. This 20\% temporal divergence is pivotal, offering healthcare systems a crucial extended window for preparation.

In conclusion, for every dollar invested, contact tracing proves to be the more cost-efficient choice compared to vaccination, particularly in the crucial first 100 days. A mere 40\% app adoption rate paired with 80\% self-quarantine compliance under the CT strategy offers a better return on investment than the same expenditure on vaccination alone.

\subsection{Coupled effect of pharmaceutical, behavioral, and digital interventions} \label{subsec:coupled-analysis}
\begin{figure*}
    \begin{center}    \includegraphics[width=0.32\textwidth,height=0.24\textwidth]{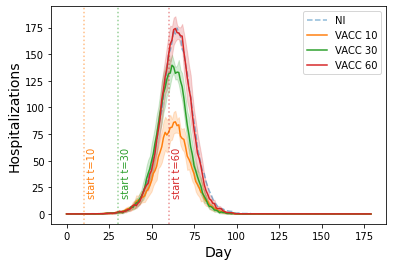}    \includegraphics[width=0.32\textwidth,height=0.24\textwidth]{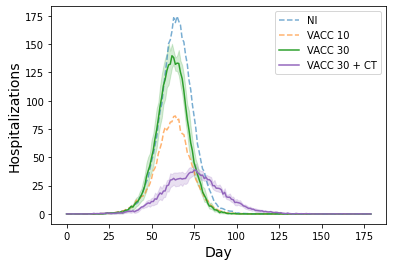}    \includegraphics[width=0.32\textwidth,height=0.24\textwidth]{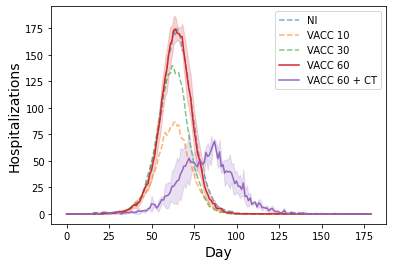}
    \caption{Analysis of interplay of digital and behavioral interventions on delayed vaccination. (a) Illustrates the impact of vaccine deployment speed on hospitalizations. Vaccine rollout delays lead to a consequential rise in hospitalizations with peak incidence remaining consistent. (b) Demonstrates the synergy of contact tracing and varied vaccine deployment timings, emphasizing that combining VACC(t = 30) + CT significantly diminishes hospitalizations and prolongs the time to peak compared to early vaccination alone. (c) Indicates the challenges with vaccine initiation at the pandemic's zenith, stressing that even late vaccine rollouts, when coupled with testing, contact tracing, and self-quarantine, can drastically mitigate infections and allow for a crucial extended immunization period. This highlights the indispensability of integrating behavioral and digital strategies, especially in the pandemic's early days when clinical interventions might not yet be in full swing.}
    \label{fig:delayed_vacc}
    \end{center}
\end{figure*}
In this section, we study the combined effects of vaccine deployment speed and other pivotal interventions, examining their collective impact on hospitalizations. We observe that regardless of intervention combinations, the peak consistently emerges within the first 100 days, highlighting the significance of timely informed decisions.  Figure \ref{fig:delayed_vacc}(a) illustrates the relationship between the speed of vaccine deployment at distinct time intervals: \( t=10 \), \( t=30 \), and \( t=60 \) and the subsequent hospitalizations. Compared to starting the vaccinations at t=10, a delayed vaccination drive starting at t=30 and t=60 increases the number of hospitalizations by 61\% and 103\%, respectively, with all scenarios peaking around the same time.

However, by integrating other digital and behavioral interventions with vaccination, we observe a transformative mitigation effect. Figure \ref{fig:delayed_vacc}(b) shows that VACC starting at t = 30 + CT leads to a 55\% reduction in hospitalizations compared to only early vaccination starting at t=10 and further a 72\% drop in hospitalizations compared with VACC at t=30. Additionally, this amalgamated approach grants an extra 14-day buffer prior to the hospitalization peak, facilitating a strategic advantage for healthcare system preparedness. The cost analysis for Figure \ref{fig:delayed_vacc}(b) is provided in supplementary material.

Further, sole reliance on late vaccination starting at \(t=60\) fails because of the inherent lag in post-inoculation immunity development. However, when this late vaccination is augmented with proactive contact tracing and self-quarantine measures, the results are noteworthy: a 61\% reduction in hospitalizations accompanied by an additional 23-day window for effective immunization as in Figure \ref{fig:delayed_vacc}(c).

This underscores the potential of non-pharmaceutical interventions (NPIs) like behavioral and digital, not just as alternate measures but as pivotal strategies in pandemic control, especially when vaccination rollout faces delays. Thus, a multifaceted approach combining behavioral, digital, and pharmaceutical measures is pivotal in effectively managing the pandemic, especially during the first 100 days when clinical interventions face delays.


\subsection{Geographical spread} \label{subsec:geographical-spread}


\begin{figure}[h!]
    \centering
    
    \includegraphics[width=0.15\textwidth,height=0.18\textwidth]{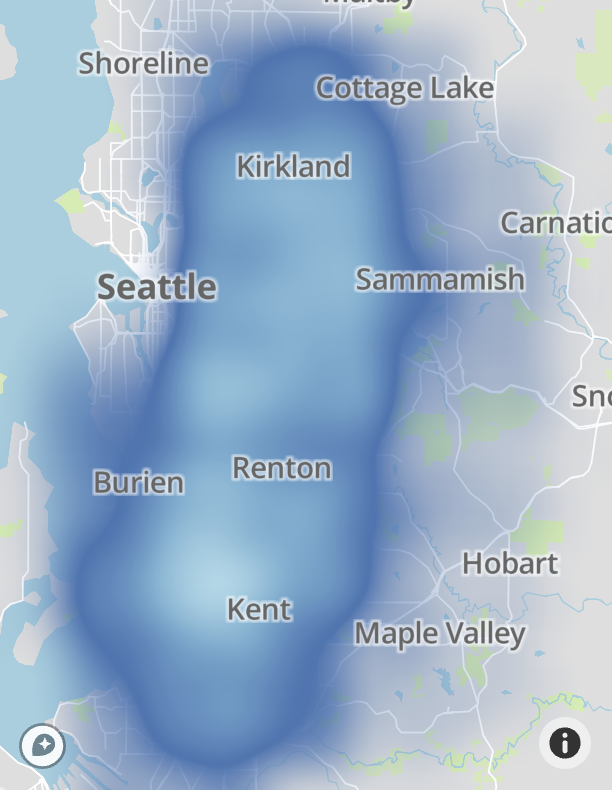}    \includegraphics[width=0.15\textwidth,height=0.18\textwidth]{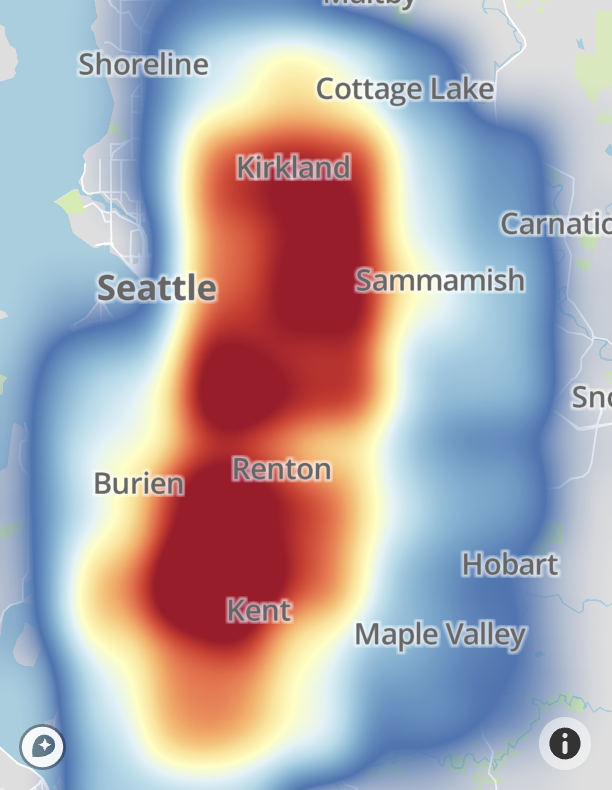}    \includegraphics[width=0.15\textwidth,height=0.18\textwidth]{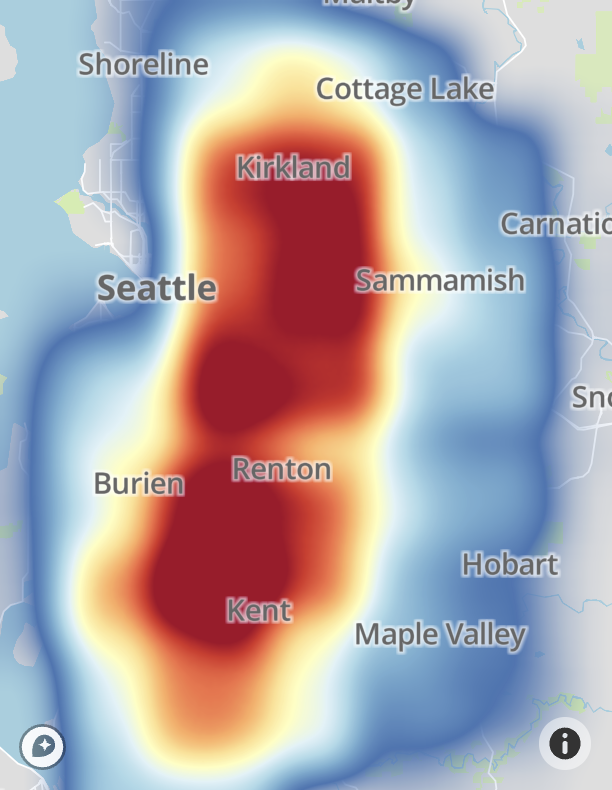}    \includegraphics[width=0.01\textwidth,height=0.18\textwidth]{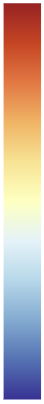}
    
    \vspace{0.2em}

   \includegraphics[width=0.15\textwidth,height=0.18\textwidth]{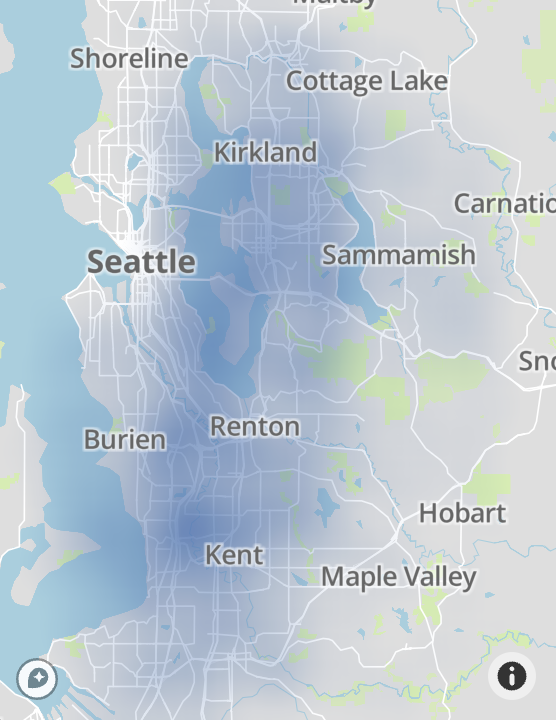}    \includegraphics[width=0.15\textwidth,height=0.18\textwidth]{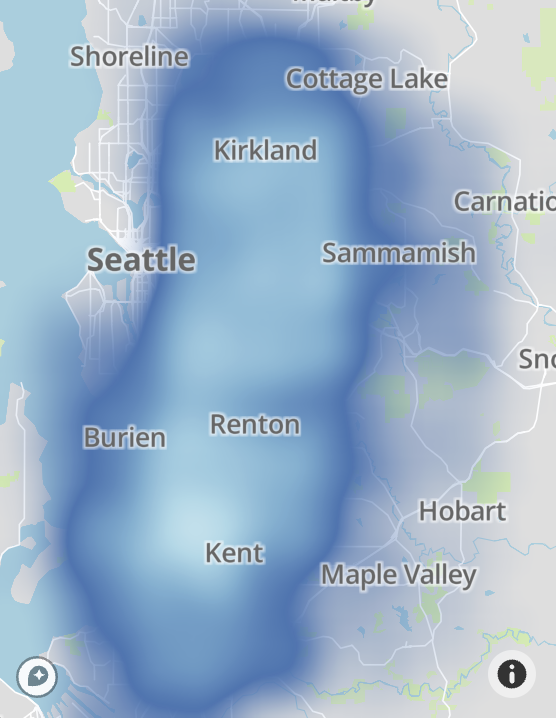}    \includegraphics[width=0.15\textwidth,height=0.18\textwidth]{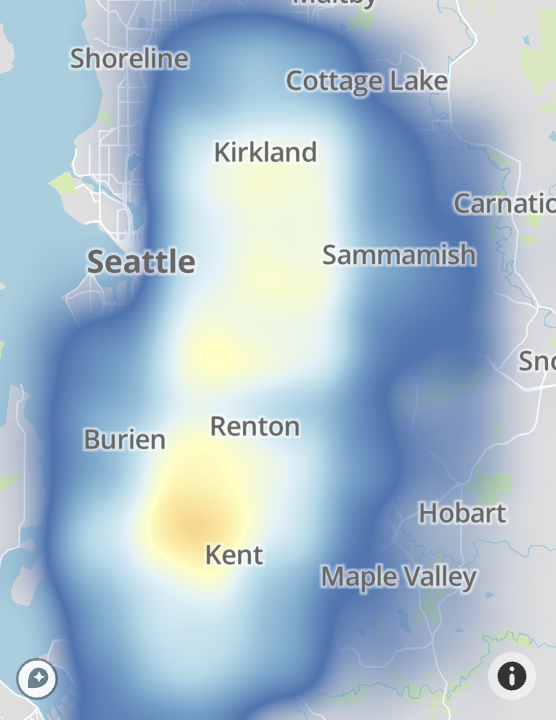}    \includegraphics[width=0.01\textwidth,height=0.18\textwidth]{figs/legend.png}

    \caption{Geographical progression of infections in Kings's County, Washington, at different time intervals. (Top) In case of no intervention, the infection spreads to 25\% of the population by t=50, 76\% by t=70, and 81\% by t=120. (Bottom) In the case of combined digital, behavioral, and pharmaceutical interventions, infection spreads slowly to only 5\% of the population by t=50, 19\% by t=70, and only 36\% by t=120}
    \label{fig:geo}
\end{figure}

Figure \ref{fig:geo} shows heat maps of infection spread over time in King's County, Washington State at distinct time intervals: \( t=50 \), \( t=70 \), and \( t=120 \). We compare the unmitigated (NI) scenario in Figure \ref{fig:geo} (Top) against an integrated strategy that combines all interventions vaccinations, testing, contact tracing, and self-quarantine in Figure \ref{fig:geo} (Bottom). We observe that in the case of no intervention, the infection spreads aggressively, already infecting a substantial 25\% of the population by day 50, 76\% by t=70, and 81\% by t=120. In stark contrast, Figure \ref{fig:geo} (Bottom) captures the attenuated spread when a comprehensive set of interventions is deployed. Infections stand at a mere 5\% population infected by day 50, 19\% by day 70, and only 36\% by t=120. 
\section{Discussion}
In this paper, we highlight the potential of agent-based models to simulate highly complex environments through multiple intertwined interventions. We discussed intricate modeling of pharmaceutical, behavioral, and digital interventions and how their holistic understanding is important when creating pandemic policies. While we take one step towards bridging the gap between understanding the emerging trends and policy-making, we posit that there can be additional implicit factors stemming from complex interventions and their ripple effects that frequently go unnoticed, yet significantly influence the trajectory of the pandemic. For instance, financial interventions like severance funds or government aid \cite{jpmorganfin} provided to the unemployed to stay at home could alter mobility patterns \cite{askitas2021estimating, alessandretti2022human}, influencing the pandemic's course. Additionally, our cost analysis focuses predominantly on explicit monetary costs. Some interventions, while not incurring direct costs, may lead to broader economic implications \cite{mckee2020if, verschuur2021global, shang2021effects}. For instance, extensive lockdowns, a common strategy during COVID-19, triggered a severe global economic downturn. This collapse was characterized by soaring unemployment rates \cite{ahmad2023impact, antipova2021analysis}, halted international trade \cite{verschuur2021observed, hayakawa2021impact, liu2022trade}, and suspended supply chains \cite{naseer2023covid, guan2020global, magableh2021supply, paul2021supply}. These influencing factors further raise pivotal questions for policy-making: Does the amount of unemployment aid play a larger role in pandemic control than the speed of its provision? Are short-term lockdowns (<31 days) the solution \cite{yamaka2022analysis, oraby2021modeling, bisiacco2021covid}, or is there an optimal percentage of people returning to offices (RTO) \cite{yang2022effects, garzillo2022returning} that can help control the pandemic and also not harm the economy at a global scale?

Answering these questions can provide essential insights into the efficacy and promptness of policy interventions. Therefore, modeling these latent factors is a vital future direction in aiming for a comprehensive understanding of a pandemic's broader impacts. This will guide us toward more nuanced policy-making for future outbreaks, ensuring both public health and economic stability.

\section{Conclusion}
In this paper, we emphasize the capabilities of agent-based models in understanding the complex dynamics of pandemics and simulating the potential impact of different policy interventions. By simulating interventions with their real-world deployment challenges, we analyze emergent behaviors on populations at scale. Our approach goes beyond merely evaluating standalone interventions by capturing the comprehensive interplay of combined strategies. From our experiments, several critical findings emerged. The initial 100 days of a pandemic largely shape its course and underline the need for swift and informed decisions from the beginning. While vaccines play a pivotal role in reducing individual susceptibility, achieving community-wide immunity is a gradual process due to the time-consuming nature of mass vaccination rollouts. Our research emphatically highlights the indispensability of sustained interventions alongside vaccinations. Notably, we observed contact tracing's efficacy for not only reducing the cumulative infections from 81\% (in the absence of intervention) to 54\% but also delaying the infection peak by 14 days. Our analysis further shows that the same amount of dollars spent on extensive testing with contact tracing and self-quarantine proves to be more cost-effective than spending on vaccinations alone. Future global health crises thus necessitate a balanced, multi-pronged response. 
\section{Supplementary material}
\subsection{More details on the simulation framework} 
Transition Mechanics: Our ABM's transition mechanics dictate the chronological progression of every agent's state variables in the system, contingent upon the agents' actions. Among these variables, the evolution related to testing, self-quarantine, and vaccination is driven by individual agent actions. However, the progression of the agent's stage is nuanced and draws from the model described in~\cite{hinch2021openabm}. The stage dynamics, which are modified SEIR dynamics at the agent level, can be divided into two parts: 
(i) Infection dynamics: This segment addresses the state transition from susceptibility to either the asymptomatic or the pre-symptomatic infectious states. The process aggregates the outcomes of interactions with all other agents, factoring in the quarantine and vaccination states of these agents. This is a Markovian transition each phase presents a probability wherein an agent transitions from a non-infected susceptible state to an infectious one. The potency of each interaction varies based on factors like age, disease phase, quarantine, vaccination status, and the facilitating interaction network. The likelihood of infection results from aggregating the influencing factors from every interaction.
(ii) Disease dynamics: This section navigates through the various infectious disease stages, culminating in either recovery or a death-absorbing state. These shifts are semi-Markov in nature, where durations are inferred from continuous-time distributions, influenced by disease stage and age parameters. 

\subsection{Cost analysis of delayed vaccine}
\begin{figure}[!h]
    \begin{center}
    \includegraphics[width=0.4\textwidth, height=0.3\textwidth]{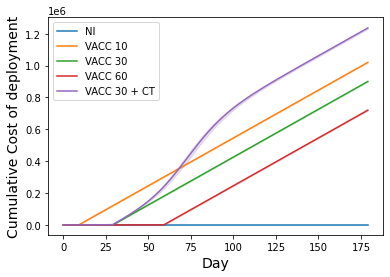}
    \caption{Cost analysis of behavioral and digital interventions with delayed vaccination (VACC+CT). Further explained in Section 4.4}
    \label{fig:cost_delayed_vaccine}
    \end{center}
\end{figure}

Figure \ref{fig:cost_delayed_vaccine} shows the cost analysis for different interventions including vaccination starting at times t=10, 30, 60, and combined intervention of vaccination at t=30 with contact tracing. For each case, once the vaccination rollout begins, 0.3\% of the population is inoculated daily. The total cost for VACC 30 + CT includes the cost of vaccines and tests for contact tracing, which is higher than the cost of administering vaccines alone starting at t=10, t=30, or t=60. More details on the impact of each of these interventions are covered in Section 4.4.

\subsection{Sensitivity Analysis}
\begin{figure*}
    \begin{center}    \includegraphics[width=0.32\textwidth,height=0.25\textwidth]{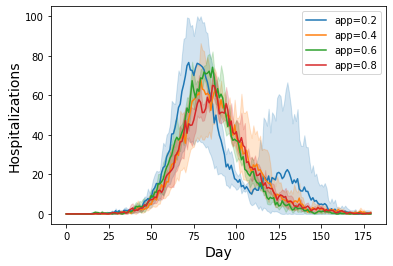}    \includegraphics[width=0.32\textwidth,height=0.25\textwidth]{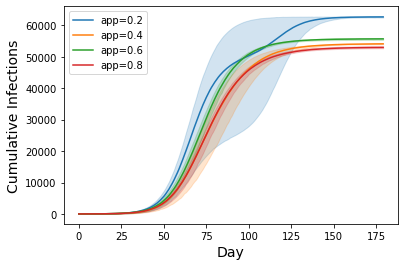}    \includegraphics[width=0.32\textwidth,height=0.25\textwidth]{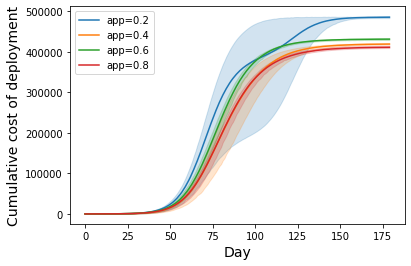}
    \caption{Sensitivity analysis of the impact of app adoption in contact tracing on hospitalizations, cumulative infections, and the cost of deployment for pandemic containment}
    \label{fig:sensitivity_app_prob}
    \end{center}
\end{figure*}

\begin{figure*}
    \begin{center}    \includegraphics[width=0.32\textwidth,height=0.25\textwidth]{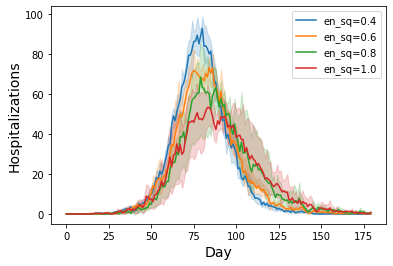}    \includegraphics[width=0.32\textwidth,height=0.25\textwidth]{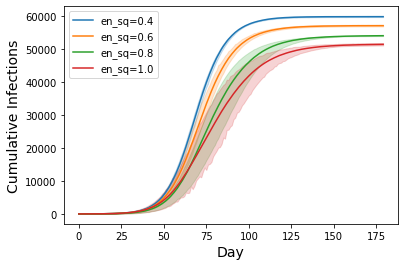}    \includegraphics[width=0.32\textwidth,height=0.25\textwidth]{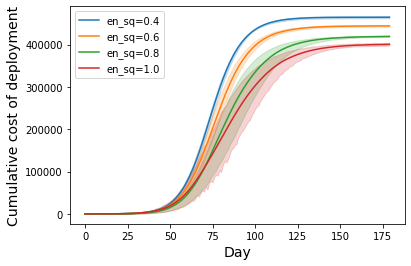}
    \caption{Sensitivity analysis of self-quarantine compliance probability in contact tracing on hospitalizations, cumulative infections, and the cost of deployment for pandemic containment}
    \label{fig:sensitivity_sq_prob}
    \end{center}
\end{figure*}

\begin{figure*}
    \begin{center}    \includegraphics[width=0.32\textwidth,height=0.25\textwidth]{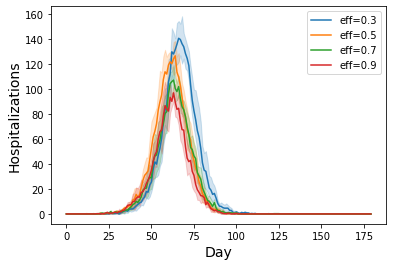}    \includegraphics[width=0.32\textwidth,height=0.25\textwidth]{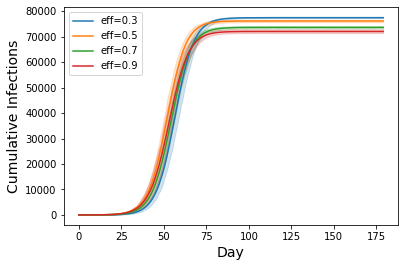}    \includegraphics[width=0.32\textwidth,height=0.25\textwidth]{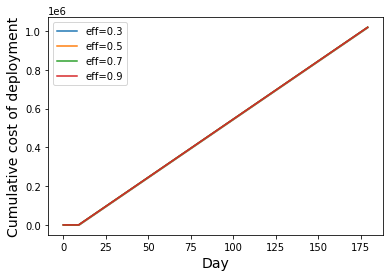}
    \caption{Sensitivity analysis of the efficacy of the first dose of vaccination on hospitalizations, cumulative infections, and the cost of deployment for pandemic containment}
    \label{fig:sensitivity_vacc_eff1}
    \end{center}
\end{figure*}

We also perform sensitivity analysis of some parameters including app adoption rate, self-quarantine compliance probability on receiving exposure notification in contact tracing, and efficacy of the first dose of vaccination on hospitalizations, cumulative infections, and cost of containing the pandemic as shown in Figures \ref{fig:sensitivity_app_prob}, \ref{fig:sensitivity_sq_prob}, and \ref{fig:sensitivity_vacc_eff1}.

\bibliographystyle{ACM-Reference-Format} 
\bibliography{ref}


\begin{thebibliography}{68}


\ifx \showCODEN    \undefined \def \showCODEN     #1{\unskip}     \fi
\ifx \showDOI      \undefined \def \showDOI       #1{#1}\fi
\ifx \showISBNx    \undefined \def \showISBNx     #1{\unskip}     \fi
\ifx \showISBNxiii \undefined \def \showISBNxiii  #1{\unskip}     \fi
\ifx \showISSN     \undefined \def \showISSN      #1{\unskip}     \fi
\ifx \showLCCN     \undefined \def \showLCCN      #1{\unskip}     \fi
\ifx \shownote     \undefined \def \shownote      #1{#1}          \fi
\ifx \showarticletitle \undefined \def \showarticletitle #1{#1}   \fi
\ifx \showURL      \undefined \def \showURL       {\relax}        \fi
\providecommand\bibfield[2]{#2}
\providecommand\bibinfo[2]{#2}
\providecommand\natexlab[1]{#1}
\providecommand\showeprint[2][]{arXiv:#2}

\bibitem[cov({[n.\,d.]})]%
        {covidbedsperc}
 \bibinfo{year}{[n.\,d.]}\natexlab{}.
\newblock \bibinfo{booktitle}{\emph{Does King County have enough hospital beds to deal with the coronavirus?}}
\newblock
\urldef\tempurl%
\url{https://sccinsight.com/2020/03/10/does-king-county-have-enough-hospital-beds-to-deal-with-the-coronavirus/}
\showURL{%
\tempurl}


\bibitem[Abueg et~al\mbox{.}(2021)]%
        {abueg2021modeling}
\bibfield{author}{\bibinfo{person}{Matthew Abueg}, \bibinfo{person}{Robert Hinch}, \bibinfo{person}{Neo Wu}, \bibinfo{person}{Luyang Liu}, \bibinfo{person}{William Probert}, \bibinfo{person}{Austin Wu}, \bibinfo{person}{Paul Eastham}, \bibinfo{person}{Yusef Shafi}, \bibinfo{person}{Matt Rosencrantz}, \bibinfo{person}{Michael Dikovsky}, {et~al\mbox{.}}} \bibinfo{year}{2021}\natexlab{}.
\newblock \showarticletitle{Modeling the effect of exposure notification and non-pharmaceutical interventions on COVID-19 transmission in Washington state}.
\newblock \bibinfo{journal}{\emph{NPJ digital medicine}} \bibinfo{volume}{4}, \bibinfo{number}{1} (\bibinfo{year}{2021}), \bibinfo{pages}{49}.
\newblock


\bibitem[Ahmad et~al\mbox{.}(2023)]%
        {ahmad2023impact}
\bibfield{author}{\bibinfo{person}{Muneeb Ahmad}, \bibinfo{person}{Yousaf~Ali Khan}, \bibinfo{person}{Chonghui Jiang}, \bibinfo{person}{Syed Jawad~Haider Kazmi}, {and} \bibinfo{person}{Syed~Zaheer Abbas}.} \bibinfo{year}{2023}\natexlab{}.
\newblock \showarticletitle{The impact of COVID-19 on unemployment rate: An intelligent based unemployment rate prediction in selected countries of Europe}.
\newblock \bibinfo{journal}{\emph{International Journal of Finance \& Economics}} \bibinfo{volume}{28}, \bibinfo{number}{1} (\bibinfo{year}{2023}), \bibinfo{pages}{528--543}.
\newblock


\bibitem[Alessandretti(2022)]%
        {alessandretti2022human}
\bibfield{author}{\bibinfo{person}{Laura Alessandretti}.} \bibinfo{year}{2022}\natexlab{}.
\newblock \showarticletitle{What human mobility data tell us about COVID-19 spread}.
\newblock \bibinfo{journal}{\emph{Nature Reviews Physics}} \bibinfo{volume}{4}, \bibinfo{number}{1} (\bibinfo{year}{2022}), \bibinfo{pages}{12--13}.
\newblock


\bibitem[Aleta et~al\mbox{.}(2020)]%
        {aleta2020modelling}
\bibfield{author}{\bibinfo{person}{Alberto Aleta}, \bibinfo{person}{David Martin-Corral}, \bibinfo{person}{Ana Pastore~y Piontti}, \bibinfo{person}{Marco Ajelli}, \bibinfo{person}{Maria Litvinova}, \bibinfo{person}{Matteo Chinazzi}, \bibinfo{person}{Natalie~E Dean}, \bibinfo{person}{M~Elizabeth Halloran}, \bibinfo{person}{Ira~M Longini~Jr}, \bibinfo{person}{Stefano Merler}, {et~al\mbox{.}}} \bibinfo{year}{2020}\natexlab{}.
\newblock \showarticletitle{Modelling the impact of testing, contact tracing and household quarantine on second waves of COVID-19}.
\newblock \bibinfo{journal}{\emph{Nature Human Behaviour}} \bibinfo{volume}{4}, \bibinfo{number}{9} (\bibinfo{year}{2020}), \bibinfo{pages}{964--971}.
\newblock


\bibitem[Alsalem et~al\mbox{.}(2022)]%
        {alsalem2022multi}
\bibfield{author}{\bibinfo{person}{MA Alsalem}, \bibinfo{person}{AH Alamoodi}, \bibinfo{person}{OS Albahri}, \bibinfo{person}{KA Dawood}, \bibinfo{person}{RT Mohammed}, \bibinfo{person}{Alhamzah Alnoor}, \bibinfo{person}{AA Zaidan}, \bibinfo{person}{AS Albahri}, \bibinfo{person}{BB Zaidan}, \bibinfo{person}{FM Jumaah}, {et~al\mbox{.}}} \bibinfo{year}{2022}\natexlab{}.
\newblock \showarticletitle{Multi-criteria decision-making for coronavirus disease 2019 applications: a theoretical analysis review}.
\newblock \bibinfo{journal}{\emph{Artificial Intelligence Review}} \bibinfo{volume}{55}, \bibinfo{number}{6} (\bibinfo{year}{2022}), \bibinfo{pages}{4979--5062}.
\newblock


\bibitem[Amzat et~al\mbox{.}(2020)]%
        {amzat2020coronavirus}
\bibfield{author}{\bibinfo{person}{Jimoh Amzat}, \bibinfo{person}{Kafayat Aminu}, \bibinfo{person}{Victor~I Kolo}, \bibinfo{person}{Ayodele~A Akinyele}, \bibinfo{person}{Janet~A Ogundairo}, {and} \bibinfo{person}{Maryann~C Danjibo}.} \bibinfo{year}{2020}\natexlab{}.
\newblock \showarticletitle{Coronavirus outbreak in Nigeria: Burden and socio-medical response during the first 100 days}.
\newblock \bibinfo{journal}{\emph{International Journal of Infectious Diseases}}  \bibinfo{volume}{98} (\bibinfo{year}{2020}), \bibinfo{pages}{218--224}.
\newblock


\bibitem[Angeli and Montefusco(2020)]%
        {angeli2020sensemaking}
\bibfield{author}{\bibinfo{person}{Federica Angeli} {and} \bibinfo{person}{Andrea Montefusco}.} \bibinfo{year}{2020}\natexlab{}.
\newblock \showarticletitle{Sensemaking and learning during the Covid-19 pandemic: A complex adaptive systems perspective on policy decision-making}.
\newblock \bibinfo{journal}{\emph{World Development}}  \bibinfo{volume}{136} (\bibinfo{year}{2020}), \bibinfo{pages}{105106}.
\newblock


\bibitem[Antipova(2021)]%
        {antipova2021analysis}
\bibfield{author}{\bibinfo{person}{Anzhelika Antipova}.} \bibinfo{year}{2021}\natexlab{}.
\newblock \showarticletitle{Analysis of the COVID-19 impacts on employment and unemployment across the multi-dimensional social disadvantaged areas}.
\newblock \bibinfo{journal}{\emph{Social Sciences \& Humanities Open}} \bibinfo{volume}{4}, \bibinfo{number}{1} (\bibinfo{year}{2021}), \bibinfo{pages}{100224}.
\newblock


\bibitem[Askitas et~al\mbox{.}(2021)]%
        {askitas2021estimating}
\bibfield{author}{\bibinfo{person}{Nikolaos Askitas}, \bibinfo{person}{Konstantinos Tatsiramos}, {and} \bibinfo{person}{Bertrand Verheyden}.} \bibinfo{year}{2021}\natexlab{}.
\newblock \showarticletitle{Estimating worldwide effects of non-pharmaceutical interventions on COVID-19 incidence and population mobility patterns using a multiple-event study}.
\newblock \bibinfo{journal}{\emph{Scientific reports}} \bibinfo{volume}{11}, \bibinfo{number}{1} (\bibinfo{year}{2021}), \bibinfo{pages}{1972}.
\newblock


\bibitem[Aylett-Bullock et~al\mbox{.}(2021)]%
        {aylett2021june}
\bibfield{author}{\bibinfo{person}{Joseph Aylett-Bullock}, \bibinfo{person}{Carolina Cuesta-Lazaro}, \bibinfo{person}{Arnau Quera-Bofarull}, \bibinfo{person}{Miguel Icaza-Lizaola}, \bibinfo{person}{Aidan Sedgewick}, \bibinfo{person}{Henry Truong}, \bibinfo{person}{Aoife Curran}, \bibinfo{person}{Edward Elliott}, \bibinfo{person}{Tristan Caulfield}, \bibinfo{person}{Kevin Fong}, {et~al\mbox{.}}} \bibinfo{year}{2021}\natexlab{}.
\newblock \showarticletitle{June: open-source individual-based epidemiology simulation}.
\newblock \bibinfo{journal}{\emph{Royal Society open science}} \bibinfo{volume}{8}, \bibinfo{number}{7} (\bibinfo{year}{2021}), \bibinfo{pages}{210506}.
\newblock


\bibitem[Banholzer et~al\mbox{.}(2021)]%
        {banholzer2021estimating}
\bibfield{author}{\bibinfo{person}{Nicolas Banholzer}, \bibinfo{person}{Eva Van~Weenen}, \bibinfo{person}{Adrian Lison}, \bibinfo{person}{Alberto Cenedese}, \bibinfo{person}{Arne Seeliger}, \bibinfo{person}{Bernhard Kratzwald}, \bibinfo{person}{Daniel Tschernutter}, \bibinfo{person}{Joan~Puig Salles}, \bibinfo{person}{Pierluigi Bottrighi}, \bibinfo{person}{Sonja Lehtinen}, {et~al\mbox{.}}} \bibinfo{year}{2021}\natexlab{}.
\newblock \showarticletitle{Estimating the effects of non-pharmaceutical interventions on the number of new infections with COVID-19 during the first epidemic wave}.
\newblock \bibinfo{journal}{\emph{PLoS one}} \bibinfo{volume}{16}, \bibinfo{number}{6} (\bibinfo{year}{2021}), \bibinfo{pages}{e0252827}.
\newblock


\bibitem[Barrat et~al\mbox{.}(2021)]%
        {barrat2021effect}
\bibfield{author}{\bibinfo{person}{Alain Barrat}, \bibinfo{person}{Ciro Cattuto}, \bibinfo{person}{Mikko Kivel{\"a}}, \bibinfo{person}{Sune Lehmann}, {and} \bibinfo{person}{Jari Saram{\"a}ki}.} \bibinfo{year}{2021}\natexlab{}.
\newblock \showarticletitle{Effect of manual and digital contact tracing on covid-19 outbreaks: A study on empirical contact data}.
\newblock \bibinfo{journal}{\emph{Journal of the Royal Society Interface}} \bibinfo{volume}{18}, \bibinfo{number}{178} (\bibinfo{year}{2021}), \bibinfo{pages}{20201000}.
\newblock


\bibitem[Bisiacco and Pillonetto(2021)]%
        {bisiacco2021covid}
\bibfield{author}{\bibinfo{person}{Mauro Bisiacco} {and} \bibinfo{person}{Gianluigi Pillonetto}.} \bibinfo{year}{2021}\natexlab{}.
\newblock \showarticletitle{COVID-19 epidemic control using short-term lockdowns for collective gain}.
\newblock \bibinfo{journal}{\emph{Annual Reviews in Control}}  \bibinfo{volume}{52} (\bibinfo{year}{2021}), \bibinfo{pages}{573--586}.
\newblock


\bibitem[Bisset et~al\mbox{.}(2009)]%
        {bisset2009epifast}
\bibfield{author}{\bibinfo{person}{Keith~R Bisset}, \bibinfo{person}{Jiangzhuo Chen}, \bibinfo{person}{Xizhou Feng}, \bibinfo{person}{VS~Anil Kumar}, {and} \bibinfo{person}{Madhav~V Marathe}.} \bibinfo{year}{2009}\natexlab{}.
\newblock \showarticletitle{EpiFast: a fast algorithm for large scale realistic epidemic simulations on distributed memory systems}. In \bibinfo{booktitle}{\emph{Proceedings of the 23rd international conference on Supercomputing}}. \bibinfo{pages}{430--439}.
\newblock


\bibitem[Bonabeau(2002)]%
        {bonabeau2002agent}
\bibfield{author}{\bibinfo{person}{Eric Bonabeau}.} \bibinfo{year}{2002}\natexlab{}.
\newblock \showarticletitle{Agent-based modeling: Methods and techniques for simulating human systems}.
\newblock \bibinfo{journal}{\emph{Proceedings of the national academy of sciences}} \bibinfo{volume}{99}, \bibinfo{number}{suppl\_3} (\bibinfo{year}{2002}), \bibinfo{pages}{7280--7287}.
\newblock


\bibitem[Brauner et~al\mbox{.}(2021)]%
        {brauner2021inferring}
\bibfield{author}{\bibinfo{person}{Jan~M Brauner}, \bibinfo{person}{S{\"o}ren Mindermann}, \bibinfo{person}{Mrinank Sharma}, \bibinfo{person}{David Johnston}, \bibinfo{person}{John Salvatier}, \bibinfo{person}{Tom{\'a}{\v{s}} Gaven{\v{c}}iak}, \bibinfo{person}{Anna~B Stephenson}, \bibinfo{person}{Gavin Leech}, \bibinfo{person}{George Altman}, \bibinfo{person}{Vladimir Mikulik}, {et~al\mbox{.}}} \bibinfo{year}{2021}\natexlab{}.
\newblock \showarticletitle{Inferring the effectiveness of government interventions against COVID-19}.
\newblock \bibinfo{journal}{\emph{Science}} \bibinfo{volume}{371}, \bibinfo{number}{6531} (\bibinfo{year}{2021}), \bibinfo{pages}{eabd9338}.
\newblock


\bibitem[Chang et~al\mbox{.}(2021)]%
        {chang2021mobility}
\bibfield{author}{\bibinfo{person}{Serina Chang}, \bibinfo{person}{Emma Pierson}, \bibinfo{person}{Pang~Wei Koh}, \bibinfo{person}{Jaline Gerardin}, \bibinfo{person}{Beth Redbird}, \bibinfo{person}{David Grusky}, {and} \bibinfo{person}{Jure Leskovec}.} \bibinfo{year}{2021}\natexlab{}.
\newblock \showarticletitle{Mobility network models of COVID-19 explain inequities and inform reopening}.
\newblock \bibinfo{journal}{\emph{Nature}} \bibinfo{volume}{589}, \bibinfo{number}{7840} (\bibinfo{year}{2021}), \bibinfo{pages}{82--87}.
\newblock


\bibitem[Cheng et~al\mbox{.}(2020)]%
        {cheng2020contact}
\bibfield{author}{\bibinfo{person}{Hao-Yuan Cheng}, \bibinfo{person}{Shu-Wan Jian}, \bibinfo{person}{Ding-Ping Liu}, \bibinfo{person}{Ta-Chou Ng}, \bibinfo{person}{Wan-Ting Huang}, \bibinfo{person}{Hsien-Ho Lin}, {et~al\mbox{.}}} \bibinfo{year}{2020}\natexlab{}.
\newblock \showarticletitle{Contact tracing assessment of COVID-19 transmission dynamics in Taiwan and risk at different exposure periods before and after symptom onset}.
\newblock \bibinfo{journal}{\emph{JAMA internal medicine}} \bibinfo{volume}{180}, \bibinfo{number}{9} (\bibinfo{year}{2020}), \bibinfo{pages}{1156--1163}.
\newblock


\bibitem[Chopra et~al\mbox{.}(2021)]%
        {chopra2021deepabm}
\bibfield{author}{\bibinfo{person}{Ayush Chopra}, \bibinfo{person}{Esma Gel}, \bibinfo{person}{Jayakumar Subramanian}, \bibinfo{person}{Balaji Krishnamurthy}, \bibinfo{person}{Santiago Romero-Brufau}, \bibinfo{person}{Kalyan~S Pasupathy}, \bibinfo{person}{Thomas~C Kingsley}, {and} \bibinfo{person}{Ramesh Raskar}.} \bibinfo{year}{2021}\natexlab{}.
\newblock \showarticletitle{DeepABM: scalable, efficient and differentiable agent-based simulations via graph neural networks}.
\newblock \bibinfo{journal}{\emph{arXiv preprint arXiv:2110.04421}} (\bibinfo{year}{2021}).
\newblock


\bibitem[Chowdhury et~al\mbox{.}(2020)]%
        {chowdhury2020covid}
\bibfield{author}{\bibinfo{person}{Mohammad Jabed~Morshed Chowdhury}, \bibinfo{person}{Md~Sadek Ferdous}, \bibinfo{person}{Kamanashis Biswas}, \bibinfo{person}{Niaz Chowdhury}, {and} \bibinfo{person}{Vallipuram Muthukkumarasamy}.} \bibinfo{year}{2020}\natexlab{}.
\newblock \showarticletitle{COVID-19 contact tracing: challenges and future directions}.
\newblock \bibinfo{journal}{\emph{Ieee Access}}  \bibinfo{volume}{8} (\bibinfo{year}{2020}), \bibinfo{pages}{225703--225729}.
\newblock


\bibitem[Davis et~al\mbox{.}(2021)]%
        {davis2021contact}
\bibfield{author}{\bibinfo{person}{Emma~L Davis}, \bibinfo{person}{Tim~CD Lucas}, \bibinfo{person}{Anna Borlase}, \bibinfo{person}{Timothy~M Pollington}, \bibinfo{person}{Sam Abbott}, \bibinfo{person}{Diepreye Ayabina}, \bibinfo{person}{Thomas Crellen}, \bibinfo{person}{Joel Hellewell}, \bibinfo{person}{Li Pi}, {et~al\mbox{.}}} \bibinfo{year}{2021}\natexlab{}.
\newblock \showarticletitle{Contact tracing is an imperfect tool for controlling COVID-19 transmission and relies on population adherence}.
\newblock \bibinfo{journal}{\emph{Nature communications}} \bibinfo{volume}{12}, \bibinfo{number}{1} (\bibinfo{year}{2021}), \bibinfo{pages}{5412}.
\newblock


\bibitem[Dimitrov and Meyers(2010)]%
        {dimitrov2010mathematical}
\bibfield{author}{\bibinfo{person}{Nedialko~B Dimitrov} {and} \bibinfo{person}{Lauren~Ancel Meyers}.} \bibinfo{year}{2010}\natexlab{}.
\newblock \showarticletitle{Mathematical approaches to infectious disease prediction and control}.
\newblock In \bibinfo{booktitle}{\emph{Risk and optimization in an uncertain world}}. \bibinfo{publisher}{INFORMS}, \bibinfo{pages}{1--25}.
\newblock


\bibitem[Du et~al\mbox{.}(2021)]%
        {du2021comparative}
\bibfield{author}{\bibinfo{person}{Zhanwei Du}, \bibinfo{person}{Abhishek Pandey}, \bibinfo{person}{Yuan Bai}, \bibinfo{person}{Meagan~C Fitzpatrick}, \bibinfo{person}{Matteo Chinazzi}, \bibinfo{person}{Ana~Pastore y Piontti}, \bibinfo{person}{Michael Lachmann}, \bibinfo{person}{Alessandro Vespignani}, \bibinfo{person}{Benjamin~J Cowling}, \bibinfo{person}{Alison~P Galvani}, {et~al\mbox{.}}} \bibinfo{year}{2021}\natexlab{}.
\newblock \showarticletitle{Comparative cost-effectiveness of SARS-CoV-2 testing strategies in the USA: a modelling study}.
\newblock \bibinfo{journal}{\emph{The Lancet Public Health}} \bibinfo{volume}{6}, \bibinfo{number}{3} (\bibinfo{year}{2021}), \bibinfo{pages}{e184--e191}.
\newblock


\bibitem[Flaxman et~al\mbox{.}(2020)]%
        {flaxman2020estimating}
\bibfield{author}{\bibinfo{person}{Seth Flaxman}, \bibinfo{person}{Swapnil Mishra}, \bibinfo{person}{Axel Gandy}, \bibinfo{person}{H~Juliette~T Unwin}, \bibinfo{person}{Thomas~A Mellan}, \bibinfo{person}{Helen Coupland}, \bibinfo{person}{Charles Whittaker}, \bibinfo{person}{Harrison Zhu}, \bibinfo{person}{Tresnia Berah}, \bibinfo{person}{Jeffrey~W Eaton}, {et~al\mbox{.}}} \bibinfo{year}{2020}\natexlab{}.
\newblock \showarticletitle{Estimating the effects of non-pharmaceutical interventions on COVID-19 in Europe}.
\newblock \bibinfo{journal}{\emph{Nature}} \bibinfo{volume}{584}, \bibinfo{number}{7820} (\bibinfo{year}{2020}), \bibinfo{pages}{257--261}.
\newblock


\bibitem[Garzillo et~al\mbox{.}(2022)]%
        {garzillo2022returning}
\bibfield{author}{\bibinfo{person}{Elpidio~Maria Garzillo}, \bibinfo{person}{Arcangelo Cioffi}, \bibinfo{person}{Angela Carta}, {and} \bibinfo{person}{Maria Grazia~Lourdes Monaco}.} \bibinfo{year}{2022}\natexlab{}.
\newblock \showarticletitle{Returning to work after the COVID-19 pandemic earthquake: a systematic review}.
\newblock \bibinfo{journal}{\emph{International journal of environmental research and public health}} \bibinfo{volume}{19}, \bibinfo{number}{8} (\bibinfo{year}{2022}), \bibinfo{pages}{4538}.
\newblock


\bibitem[Groff et~al\mbox{.}(2019)]%
        {groff2019state}
\bibfield{author}{\bibinfo{person}{Elizabeth~R Groff}, \bibinfo{person}{Shane~D Johnson}, {and} \bibinfo{person}{Amy Thornton}.} \bibinfo{year}{2019}\natexlab{}.
\newblock \showarticletitle{State of the art in agent-based modeling of urban crime: An overview}.
\newblock \bibinfo{journal}{\emph{Journal of Quantitative Criminology}}  \bibinfo{volume}{35} (\bibinfo{year}{2019}), \bibinfo{pages}{155--193}.
\newblock


\bibitem[Guan et~al\mbox{.}(2020)]%
        {guan2020global}
\bibfield{author}{\bibinfo{person}{Dabo Guan}, \bibinfo{person}{Daoping Wang}, \bibinfo{person}{Stephane Hallegatte}, \bibinfo{person}{Steven~J Davis}, \bibinfo{person}{Jingwen Huo}, \bibinfo{person}{Shuping Li}, \bibinfo{person}{Yangchun Bai}, \bibinfo{person}{Tianyang Lei}, \bibinfo{person}{Qianyu Xue}, \bibinfo{person}{D’Maris Coffman}, {et~al\mbox{.}}} \bibinfo{year}{2020}\natexlab{}.
\newblock \showarticletitle{Global supply-chain effects of COVID-19 control measures}.
\newblock \bibinfo{journal}{\emph{Nature human behaviour}} \bibinfo{volume}{4}, \bibinfo{number}{6} (\bibinfo{year}{2020}), \bibinfo{pages}{577--587}.
\newblock


\bibitem[Haug et~al\mbox{.}(2020)]%
        {haug2020ranking}
\bibfield{author}{\bibinfo{person}{Nina Haug}, \bibinfo{person}{Lukas Geyrhofer}, \bibinfo{person}{Alessandro Londei}, \bibinfo{person}{Elma Dervic}, \bibinfo{person}{Am{\'e}lie Desvars-Larrive}, \bibinfo{person}{Vittorio Loreto}, \bibinfo{person}{Beate Pinior}, \bibinfo{person}{Stefan Thurner}, {and} \bibinfo{person}{Peter Klimek}.} \bibinfo{year}{2020}\natexlab{}.
\newblock \showarticletitle{Ranking the effectiveness of worldwide COVID-19 government interventions}.
\newblock \bibinfo{journal}{\emph{Nature human behaviour}} \bibinfo{volume}{4}, \bibinfo{number}{12} (\bibinfo{year}{2020}), \bibinfo{pages}{1303--1312}.
\newblock


\bibitem[Hayakawa and Mukunoki(2021)]%
        {hayakawa2021impact}
\bibfield{author}{\bibinfo{person}{Kazunobu Hayakawa} {and} \bibinfo{person}{Hiroshi Mukunoki}.} \bibinfo{year}{2021}\natexlab{}.
\newblock \showarticletitle{The impact of COVID-19 on international trade: Evidence from the first shock}.
\newblock \bibinfo{journal}{\emph{Journal of the Japanese and International Economies}}  \bibinfo{volume}{60} (\bibinfo{year}{2021}), \bibinfo{pages}{101135}.
\newblock


\bibitem[Hethcote(2000)]%
        {hethcote2000mathematics}
\bibfield{author}{\bibinfo{person}{Herbert~W Hethcote}.} \bibinfo{year}{2000}\natexlab{}.
\newblock \showarticletitle{The mathematics of infectious diseases}.
\newblock \bibinfo{journal}{\emph{SIAM review}} \bibinfo{volume}{42}, \bibinfo{number}{4} (\bibinfo{year}{2000}), \bibinfo{pages}{599--653}.
\newblock


\bibitem[Hinch et~al\mbox{.}(2021)]%
        {hinch2021openabm}
\bibfield{author}{\bibinfo{person}{Robert Hinch}, \bibinfo{person}{William~JM Probert}, \bibinfo{person}{Anel Nurtay}, \bibinfo{person}{Michelle Kendall}, \bibinfo{person}{Chris Wymant}, \bibinfo{person}{Matthew Hall}, \bibinfo{person}{Katrina Lythgoe}, \bibinfo{person}{Ana Bulas~Cruz}, \bibinfo{person}{Lele Zhao}, \bibinfo{person}{Andrea Stewart}, {et~al\mbox{.}}} \bibinfo{year}{2021}\natexlab{}.
\newblock \showarticletitle{OpenABM-Covid19—An agent-based model for non-pharmaceutical interventions against COVID-19 including contact tracing}.
\newblock \bibinfo{journal}{\emph{PLoS computational biology}} \bibinfo{volume}{17}, \bibinfo{number}{7} (\bibinfo{year}{2021}), \bibinfo{pages}{e1009146}.
\newblock


\bibitem[Institute({[n.\,d.]})]%
        {jpmorganfin}
\bibfield{author}{\bibinfo{person}{JPMorgan~Chase Institute}.} \bibinfo{year}{[n.\,d.]}\natexlab{}.
\newblock \bibinfo{booktitle}{\emph{The First 100 Days and Beyond}}.
\newblock
\urldef\tempurl%
\url{https://www.jpmorganchase.com/content/dam/jpmc/jpmorgan-chase-and-co/institute/pdf/institute-policycenter-first-100-days_vF.pdf}
\showURL{%
\tempurl}


\bibitem[Kates et~al\mbox{.}(2023)]%
        {kates2023much}
\bibfield{author}{\bibinfo{person}{Jennifer Kates}, \bibinfo{person}{Cynthia Cox}, {and} \bibinfo{person}{Josh Michaud}.} \bibinfo{year}{2023}\natexlab{}.
\newblock \showarticletitle{How much could COVID-19 Vaccines cost the US after Commercialization}.
\newblock \bibinfo{journal}{\emph{KFF. March}}  \bibinfo{volume}{10} (\bibinfo{year}{2023}).
\newblock


\bibitem[Kerr et~al\mbox{.}(2021)]%
        {kerr2021covasim}
\bibfield{author}{\bibinfo{person}{Cliff~C Kerr}, \bibinfo{person}{Robyn~M Stuart}, \bibinfo{person}{Dina Mistry}, \bibinfo{person}{Romesh~G Abeysuriya}, \bibinfo{person}{Katherine Rosenfeld}, \bibinfo{person}{Gregory~R Hart}, \bibinfo{person}{Rafael~C N{\'u}{\~n}ez}, \bibinfo{person}{Jamie~A Cohen}, \bibinfo{person}{Prashanth Selvaraj}, \bibinfo{person}{Brittany Hagedorn}, {et~al\mbox{.}}} \bibinfo{year}{2021}\natexlab{}.
\newblock \showarticletitle{Covasim: an agent-based model of COVID-19 dynamics and interventions}.
\newblock \bibinfo{journal}{\emph{PLOS Computational Biology}} \bibinfo{volume}{17}, \bibinfo{number}{7} (\bibinfo{year}{2021}), \bibinfo{pages}{e1009149}.
\newblock


\bibitem[Kostoulas et~al\mbox{.}(2021)]%
        {kostoulas2021diagnostic}
\bibfield{author}{\bibinfo{person}{Polychronis Kostoulas}, \bibinfo{person}{Paolo Eusebi}, {and} \bibinfo{person}{Sonja Hartnack}.} \bibinfo{year}{2021}\natexlab{}.
\newblock \showarticletitle{Diagnostic accuracy estimates for COVID-19 real-time polymerase chain reaction and lateral flow immunoassay tests with bayesian latent-class models}.
\newblock \bibinfo{journal}{\emph{American journal of epidemiology}} \bibinfo{volume}{190}, \bibinfo{number}{8} (\bibinfo{year}{2021}), \bibinfo{pages}{1689--1695}.
\newblock


\bibitem[Larremore et~al\mbox{.}(2021)]%
        {larremore2021test}
\bibfield{author}{\bibinfo{person}{Daniel~B Larremore}, \bibinfo{person}{Bryan Wilder}, \bibinfo{person}{Evan Lester}, \bibinfo{person}{Soraya Shehata}, \bibinfo{person}{James~M Burke}, \bibinfo{person}{James~A Hay}, \bibinfo{person}{Milind Tambe}, \bibinfo{person}{Michael~J Mina}, {and} \bibinfo{person}{Roy Parker}.} \bibinfo{year}{2021}\natexlab{}.
\newblock \showarticletitle{Test sensitivity is secondary to frequency and turnaround time for COVID-19 screening}.
\newblock \bibinfo{journal}{\emph{Science advances}} \bibinfo{volume}{7}, \bibinfo{number}{1} (\bibinfo{year}{2021}), \bibinfo{pages}{eabd5393}.
\newblock


\bibitem[Lewis et~al\mbox{.}(2022)]%
        {lewis2022covid}
\bibfield{author}{\bibinfo{person}{Rhonda~K Lewis}, \bibinfo{person}{Pamela~P Martin}, {and} \bibinfo{person}{Bianca~L Guzman}.} \bibinfo{year}{2022}\natexlab{}.
\newblock \bibinfo{title}{COVID-19 and vulnerable populations}.
\newblock , \bibinfo{numpages}{2537--2541}~pages.
\newblock


\bibitem[Li and Guo(2020)]%
        {li2020covid}
\bibfield{author}{\bibinfo{person}{Jinfeng Li} {and} \bibinfo{person}{Xinyi Guo}.} \bibinfo{year}{2020}\natexlab{}.
\newblock \showarticletitle{COVID-19 contact-tracing apps: A survey on the global deployment and challenges}.
\newblock \bibinfo{journal}{\emph{arXiv preprint arXiv:2005.03599}} (\bibinfo{year}{2020}).
\newblock


\bibitem[Liss and Sumida({[n.\,d.]})]%
        {covidhospcapacity}
\bibfield{author}{\bibinfo{person}{Samantha Liss} {and} \bibinfo{person}{Nami Sumida}.} \bibinfo{year}{[n.\,d.]}\natexlab{}.
\newblock \bibinfo{booktitle}{\emph{How hospital capacity varies dramatically across the country}}.
\newblock
\urldef\tempurl%
\url{https://www.healthcaredive.com/news/how-hospital-capacity-varies-dramatically-across-the-country/574892/}
\showURL{%
\tempurl}


\bibitem[Liu et~al\mbox{.}(2022)]%
        {liu2022trade}
\bibfield{author}{\bibinfo{person}{Xuepeng Liu}, \bibinfo{person}{Emanuel Ornelas}, {and} \bibinfo{person}{Huimin Shi}.} \bibinfo{year}{2022}\natexlab{}.
\newblock \showarticletitle{The trade impact of the Covid-19 pandemic}.
\newblock \bibinfo{journal}{\emph{The World Economy}} \bibinfo{volume}{45}, \bibinfo{number}{12} (\bibinfo{year}{2022}), \bibinfo{pages}{3751--3779}.
\newblock


\bibitem[Magableh(2021)]%
        {magableh2021supply}
\bibfield{author}{\bibinfo{person}{Ghazi~M Magableh}.} \bibinfo{year}{2021}\natexlab{}.
\newblock \showarticletitle{Supply chains and the COVID-19 pandemic: A comprehensive framework}.
\newblock \bibinfo{journal}{\emph{European Management Review}} \bibinfo{volume}{18}, \bibinfo{number}{3} (\bibinfo{year}{2021}), \bibinfo{pages}{363--382}.
\newblock


\bibitem[Marathe and Vullikanti(2013)]%
        {marathe2013computational}
\bibfield{author}{\bibinfo{person}{Madhav Marathe} {and} \bibinfo{person}{Anil Kumar~S Vullikanti}.} \bibinfo{year}{2013}\natexlab{}.
\newblock \showarticletitle{Computational epidemiology}.
\newblock \bibinfo{journal}{\emph{Commun. ACM}} \bibinfo{volume}{56}, \bibinfo{number}{7} (\bibinfo{year}{2013}), \bibinfo{pages}{88--96}.
\newblock


\bibitem[McKee and Stuckler(2020)]%
        {mckee2020if}
\bibfield{author}{\bibinfo{person}{Martin McKee} {and} \bibinfo{person}{David Stuckler}.} \bibinfo{year}{2020}\natexlab{}.
\newblock \showarticletitle{If the world fails to protect the economy, COVID-19 will damage health not just now but also in the future}.
\newblock \bibinfo{journal}{\emph{Nature Medicine}} \bibinfo{volume}{26}, \bibinfo{number}{5} (\bibinfo{year}{2020}), \bibinfo{pages}{640--642}.
\newblock


\bibitem[Miyah et~al\mbox{.}(2022)]%
        {miyah2022covid}
\bibfield{author}{\bibinfo{person}{Youssef Miyah}, \bibinfo{person}{Mohammed Benjelloun}, \bibinfo{person}{Sanae Lairini}, \bibinfo{person}{Anissa Lahrichi}, {et~al\mbox{.}}} \bibinfo{year}{2022}\natexlab{}.
\newblock \showarticletitle{COVID-19 impact on public health, environment, human psychology, global socioeconomy, and education}.
\newblock \bibinfo{journal}{\emph{The Scientific World Journal}}  \bibinfo{volume}{2022} (\bibinfo{year}{2022}).
\newblock


\bibitem[Naseer et~al\mbox{.}(2023)]%
        {naseer2023covid}
\bibfield{author}{\bibinfo{person}{Saira Naseer}, \bibinfo{person}{Sidra Khalid}, \bibinfo{person}{Summaira Parveen}, \bibinfo{person}{Kashif Abbass}, \bibinfo{person}{Huaming Song}, {and} \bibinfo{person}{Monica~Violeta Achim}.} \bibinfo{year}{2023}\natexlab{}.
\newblock \showarticletitle{COVID-19 outbreak: Impact on global economy}.
\newblock \bibinfo{journal}{\emph{Frontiers in public health}}  \bibinfo{volume}{10} (\bibinfo{year}{2023}), \bibinfo{pages}{1009393}.
\newblock


\bibitem[Norton et~al\mbox{.}(2019)]%
        {norton2019multiscale}
\bibfield{author}{\bibinfo{person}{Kerri-Ann Norton}, \bibinfo{person}{Chang Gong}, \bibinfo{person}{Samira Jamalian}, {and} \bibinfo{person}{Aleksander~S Popel}.} \bibinfo{year}{2019}\natexlab{}.
\newblock \showarticletitle{Multiscale agent-based and hybrid modeling of the tumor immune microenvironment}.
\newblock \bibinfo{journal}{\emph{Processes}} \bibinfo{volume}{7}, \bibinfo{number}{1} (\bibinfo{year}{2019}), \bibinfo{pages}{37}.
\newblock


\bibitem[Odone et~al\mbox{.}(2020)]%
        {odone2020vaccine}
\bibfield{author}{\bibinfo{person}{Anna Odone}, \bibinfo{person}{Daria Bucci}, \bibinfo{person}{Roberto Croci}, \bibinfo{person}{Matteo Ricc{\`o}}, \bibinfo{person}{Paola Affanni}, {and} \bibinfo{person}{Carlo Signorelli}.} \bibinfo{year}{2020}\natexlab{}.
\newblock \showarticletitle{Vaccine hesitancy in COVID-19 times. An update from Italy before flu season starts}.
\newblock \bibinfo{journal}{\emph{Acta Bio Medica: Atenei Parmensis}} \bibinfo{volume}{91}, \bibinfo{number}{3} (\bibinfo{year}{2020}), \bibinfo{pages}{e2020031}.
\newblock


\bibitem[Oraby et~al\mbox{.}(2021)]%
        {oraby2021modeling}
\bibfield{author}{\bibinfo{person}{Tamer Oraby}, \bibinfo{person}{Michael~G Tyshenko}, \bibinfo{person}{Jose~Campo Maldonado}, \bibinfo{person}{Kristina Vatcheva}, \bibinfo{person}{Susie Elsaadany}, \bibinfo{person}{Walid~Q Alali}, \bibinfo{person}{Joseph~C Longenecker}, {and} \bibinfo{person}{Mustafa Al-Zoughool}.} \bibinfo{year}{2021}\natexlab{}.
\newblock \showarticletitle{Modeling the effect of lockdown timing as a COVID-19 control measure in countries with differing social contacts}.
\newblock \bibinfo{journal}{\emph{Scientific reports}} \bibinfo{volume}{11}, \bibinfo{number}{1} (\bibinfo{year}{2021}), \bibinfo{pages}{3354}.
\newblock


\bibitem[Pandey et~al\mbox{.}(2021)]%
        {pandey2021challenges}
\bibfield{author}{\bibinfo{person}{Abhishek Pandey}, \bibinfo{person}{Pratha Sah}, \bibinfo{person}{Seyed~M Moghadas}, \bibinfo{person}{Sandip Mandal}, \bibinfo{person}{Sandip Banerjee}, \bibinfo{person}{Peter~J Hotez}, {and} \bibinfo{person}{Alison~P Galvani}.} \bibinfo{year}{2021}\natexlab{}.
\newblock \showarticletitle{Challenges facing COVID-19 vaccination in India: Lessons from the initial vaccine rollout}.
\newblock \bibinfo{journal}{\emph{Journal of Global Health}}  \bibinfo{volume}{11} (\bibinfo{year}{2021}).
\newblock


\bibitem[Paul et~al\mbox{.}(2021)]%
        {paul2021supply}
\bibfield{author}{\bibinfo{person}{Sanjoy~Kumar Paul}, \bibinfo{person}{Priyabrata Chowdhury}, \bibinfo{person}{Md~Abdul Moktadir}, {and} \bibinfo{person}{Kwok~Hung Lau}.} \bibinfo{year}{2021}\natexlab{}.
\newblock \showarticletitle{Supply chain recovery challenges in the wake of COVID-19 pandemic}.
\newblock \bibinfo{journal}{\emph{Journal of Business Research}}  \bibinfo{volume}{136} (\bibinfo{year}{2021}), \bibinfo{pages}{316--329}.
\newblock


\bibitem[Pellis et~al\mbox{.}(2015)]%
        {pellis2015eight}
\bibfield{author}{\bibinfo{person}{Lorenzo Pellis}, \bibinfo{person}{Frank Ball}, \bibinfo{person}{Shweta Bansal}, \bibinfo{person}{Ken Eames}, \bibinfo{person}{Thomas House}, \bibinfo{person}{Valerie Isham}, {and} \bibinfo{person}{Pieter Trapman}.} \bibinfo{year}{2015}\natexlab{}.
\newblock \showarticletitle{Eight challenges for network epidemic models}.
\newblock \bibinfo{journal}{\emph{Epidemics}}  \bibinfo{volume}{10} (\bibinfo{year}{2015}), \bibinfo{pages}{58--62}.
\newblock


\bibitem[Raymenants et~al\mbox{.}(2022)]%
        {raymenants2022empirical}
\bibfield{author}{\bibinfo{person}{Joren Raymenants}, \bibinfo{person}{Caspar Geenen}, \bibinfo{person}{Jonathan Thibaut}, \bibinfo{person}{Klaas Nelissen}, \bibinfo{person}{Sarah Gorissen}, {and} \bibinfo{person}{Emmanuel Andre}.} \bibinfo{year}{2022}\natexlab{}.
\newblock \showarticletitle{Empirical evidence on the efficiency of backward contact tracing in COVID-19}.
\newblock \bibinfo{journal}{\emph{Nature Communications}} \bibinfo{volume}{13}, \bibinfo{number}{1} (\bibinfo{year}{2022}), \bibinfo{pages}{4750}.
\newblock


\bibitem[Reiker et~al\mbox{.}(2021)]%
        {reiker2021machine}
\bibfield{author}{\bibinfo{person}{Theresa Reiker}, \bibinfo{person}{Monica Golumbeanu}, \bibinfo{person}{Andrew Shattock}, \bibinfo{person}{Lydia Burgert}, \bibinfo{person}{Thomas~A Smith}, \bibinfo{person}{Sarah Filippi}, \bibinfo{person}{Ewan Cameron}, {and} \bibinfo{person}{Melissa~A Penny}.} \bibinfo{year}{2021}\natexlab{}.
\newblock \showarticletitle{Machine learning approaches to calibrate individual-based infectious disease models}.
\newblock \bibinfo{journal}{\emph{medRxiv}} (\bibinfo{year}{2021}), \bibinfo{pages}{2021--01}.
\newblock


\bibitem[Romero-Brufau et~al\mbox{.}(2021)]%
        {romero2021public}
\bibfield{author}{\bibinfo{person}{Santiago Romero-Brufau}, \bibinfo{person}{Ayush Chopra}, \bibinfo{person}{Alex~J Ryu}, \bibinfo{person}{Esma Gel}, \bibinfo{person}{Ramesh Raskar}, \bibinfo{person}{Walter Kremers}, \bibinfo{person}{Karen Anderson}, \bibinfo{person}{Jayakumar Subramanian}, \bibinfo{person}{Balaji Krishnamurthy}, \bibinfo{person}{Abhishek Singh}, {et~al\mbox{.}}} \bibinfo{year}{2021}\natexlab{}.
\newblock \showarticletitle{The Public Health Impact of Delaying a Second Dose of the BNT162b2 or mRNA-1273 COVID-19 Vaccine}.
\newblock \bibinfo{journal}{\emph{medRxiv}} (\bibinfo{year}{2021}), \bibinfo{pages}{2021--02}.
\newblock


\bibitem[Self et~al\mbox{.}(2021)]%
        {self2021comparative}
\bibfield{author}{\bibinfo{person}{Wesley~H Self}, \bibinfo{person}{Mark~W Tenforde}, \bibinfo{person}{Jillian~P Rhoads}, \bibinfo{person}{Manjusha Gaglani}, \bibinfo{person}{Adit~A Ginde}, \bibinfo{person}{David~J Douin}, \bibinfo{person}{Samantha~M Olson}, \bibinfo{person}{H~Keipp Talbot}, \bibinfo{person}{Jonathan~D Casey}, \bibinfo{person}{Nicholas~M Mohr}, {et~al\mbox{.}}} \bibinfo{year}{2021}\natexlab{}.
\newblock \showarticletitle{Comparative effectiveness of Moderna, Pfizer-BioNTech, and Janssen (Johnson \& Johnson) vaccines in preventing COVID-19 hospitalizations among adults without immunocompromising conditions—United States, March--August 2021}.
\newblock \bibinfo{journal}{\emph{Morbidity and Mortality Weekly Report}} \bibinfo{volume}{70}, \bibinfo{number}{38} (\bibinfo{year}{2021}), \bibinfo{pages}{1337}.
\newblock


\bibitem[Shang et~al\mbox{.}(2021)]%
        {shang2021effects}
\bibfield{author}{\bibinfo{person}{Yunfeng Shang}, \bibinfo{person}{Haiwei Li}, {and} \bibinfo{person}{Ren Zhang}.} \bibinfo{year}{2021}\natexlab{}.
\newblock \showarticletitle{Effects of pandemic outbreak on economies: evidence from business history context}.
\newblock \bibinfo{journal}{\emph{Frontiers in public health}}  \bibinfo{volume}{9} (\bibinfo{year}{2021}), \bibinfo{pages}{146}.
\newblock


\bibitem[Subbarao(2021)]%
        {subbarao2021success}
\bibfield{author}{\bibinfo{person}{Kanta Subbarao}.} \bibinfo{year}{2021}\natexlab{}.
\newblock \showarticletitle{The success of SARS-CoV-2 vaccines and challenges ahead}.
\newblock \bibinfo{journal}{\emph{Cell host \& microbe}} \bibinfo{volume}{29}, \bibinfo{number}{7} (\bibinfo{year}{2021}), \bibinfo{pages}{1111--1123}.
\newblock


\bibitem[Tam et~al\mbox{.}(2020)]%
        {tam2020use}
\bibfield{author}{\bibinfo{person}{Derrick~Y Tam}, \bibinfo{person}{David Naimark}, \bibinfo{person}{Madhu~K Natarajan}, \bibinfo{person}{Graham Woodward}, \bibinfo{person}{Garth Oakes}, \bibinfo{person}{Mirna Rahal}, \bibinfo{person}{Kali Barrett}, \bibinfo{person}{Yasin~A Khan}, \bibinfo{person}{Raphael Ximenes}, \bibinfo{person}{Stephen Mac}, {et~al\mbox{.}}} \bibinfo{year}{2020}\natexlab{}.
\newblock \showarticletitle{The use of decision modelling to inform timely policy decisions on cardiac resource capacity during the COVID-19 pandemic}.
\newblock \bibinfo{journal}{\emph{Canadian Journal of Cardiology}} \bibinfo{volume}{36}, \bibinfo{number}{8} (\bibinfo{year}{2020}), \bibinfo{pages}{1308--1312}.
\newblock


\bibitem[Tellis et~al\mbox{.}(2020)]%
        {tellis2020price}
\bibfield{author}{\bibinfo{person}{Gerard~J Tellis}, \bibinfo{person}{Nitish Sood}, {and} \bibinfo{person}{Ashish Sood}.} \bibinfo{year}{2020}\natexlab{}.
\newblock \showarticletitle{Price of delay in Covid-19 lockdowns: Delays spike total cases, natural experiments reveal}.
\newblock \bibinfo{journal}{\emph{USC Marshall School of Business Research Paper}} (\bibinfo{year}{2020}).
\newblock


\bibitem[Verschuur et~al\mbox{.}(2021a)]%
        {verschuur2021global}
\bibfield{author}{\bibinfo{person}{Jasper Verschuur}, \bibinfo{person}{Elco~E Koks}, {and} \bibinfo{person}{Jim~W Hall}.} \bibinfo{year}{2021}\natexlab{a}.
\newblock \showarticletitle{Global economic impacts of COVID-19 lockdown measures stand out in high-frequency shipping data}.
\newblock \bibinfo{journal}{\emph{PloS one}} \bibinfo{volume}{16}, \bibinfo{number}{4} (\bibinfo{year}{2021}), \bibinfo{pages}{e0248818}.
\newblock


\bibitem[Verschuur et~al\mbox{.}(2021b)]%
        {verschuur2021observed}
\bibfield{author}{\bibinfo{person}{Jasper Verschuur}, \bibinfo{person}{Elco~E Koks}, {and} \bibinfo{person}{Jim~W Hall}.} \bibinfo{year}{2021}\natexlab{b}.
\newblock \showarticletitle{Observed impacts of the COVID-19 pandemic on global trade}.
\newblock \bibinfo{journal}{\emph{Nature Human Behaviour}} \bibinfo{volume}{5}, \bibinfo{number}{3} (\bibinfo{year}{2021}), \bibinfo{pages}{305--307}.
\newblock


\bibitem[Willem et~al\mbox{.}(2021)]%
        {willem2021impact}
\bibfield{author}{\bibinfo{person}{Lander Willem}, \bibinfo{person}{Steven Abrams}, \bibinfo{person}{Pieter~JK Libin}, \bibinfo{person}{Pietro Coletti}, \bibinfo{person}{Elise Kuylen}, \bibinfo{person}{Oana Petrof}, \bibinfo{person}{Signe M{\o}gelmose}, \bibinfo{person}{James Wambua}, \bibinfo{person}{Sereina~A Herzog}, \bibinfo{person}{Christel Faes}, {et~al\mbox{.}}} \bibinfo{year}{2021}\natexlab{}.
\newblock \showarticletitle{The impact of contact tracing and household bubbles on deconfinement strategies for COVID-19}.
\newblock \bibinfo{journal}{\emph{Nature communications}} \bibinfo{volume}{12}, \bibinfo{number}{1} (\bibinfo{year}{2021}), \bibinfo{pages}{1524}.
\newblock


\bibitem[Xiao et~al\mbox{.}(2020)]%
        {xiaopharmadiff}
\bibfield{author}{\bibinfo{person}{Y. Xiao}, \bibinfo{person}{Y. Zhang}, \bibinfo{person}{D. Geng}, \bibinfo{person}{D. Cong}, \bibinfo{person}{K. Shi}, {and} \bibinfo{person}{R.~J. Knapp}.} \bibinfo{year}{2020}\natexlab{}.
\newblock \showarticletitle{Challenges of drug development during the covid‐19 pandemic: key considerations for clinical trial designs}.
\newblock \bibinfo{journal}{\emph{British Journal of Clinical Pharmacology}}  \bibinfo{volume}{87} (\bibinfo{year}{2020}), \bibinfo{pages}{2170--2185}.
\newblock
Issue 5.
\urldef\tempurl%
\url{https://doi.org/10.1111/bcp.14629}
\showDOI{\tempurl}


\bibitem[Yamaka et~al\mbox{.}(2022)]%
        {yamaka2022analysis}
\bibfield{author}{\bibinfo{person}{Woraphon Yamaka}, \bibinfo{person}{Siritaya Lomwanawong}, \bibinfo{person}{Darin Magel}, {and} \bibinfo{person}{Paravee Maneejuk}.} \bibinfo{year}{2022}\natexlab{}.
\newblock \showarticletitle{Analysis of the Lockdown Effects on the Economy, Environment, and COVID-19 Spread: Lesson Learnt from a Global Pandemic in 2020}.
\newblock \bibinfo{journal}{\emph{International Journal of Environmental Research and Public Health}} \bibinfo{volume}{19}, \bibinfo{number}{19} (\bibinfo{year}{2022}), \bibinfo{pages}{12868}.
\newblock


\bibitem[Yang et~al\mbox{.}(2022)]%
        {yang2022effects}
\bibfield{author}{\bibinfo{person}{Longqi Yang}, \bibinfo{person}{David Holtz}, \bibinfo{person}{Sonia Jaffe}, \bibinfo{person}{Siddharth Suri}, \bibinfo{person}{Shilpi Sinha}, \bibinfo{person}{Jeffrey Weston}, \bibinfo{person}{Connor Joyce}, \bibinfo{person}{Neha Shah}, \bibinfo{person}{Kevin Sherman}, \bibinfo{person}{Brent Hecht}, {et~al\mbox{.}}} \bibinfo{year}{2022}\natexlab{}.
\newblock \showarticletitle{The effects of remote work on collaboration among information workers}.
\newblock \bibinfo{journal}{\emph{Nature human behaviour}} \bibinfo{volume}{6}, \bibinfo{number}{1} (\bibinfo{year}{2022}), \bibinfo{pages}{43--54}.
\newblock


\bibitem[Yanovskiy and Socol(2022)]%
        {moshe2022lockdowns}
\bibfield{author}{\bibinfo{person}{Moshe Yanovskiy} {and} \bibinfo{person}{Yehoshua Socol}.} \bibinfo{year}{2022}\natexlab{}.
\newblock \showarticletitle{Are Lockdowns Effective in Managing Pandemics?}
\newblock \bibinfo{journal}{\emph{International Journal of Environmental Research and Public Health}}  \bibinfo{volume}{19} (\bibinfo{date}{07} \bibinfo{year}{2022}), \bibinfo{pages}{9295}.
\newblock
\urldef\tempurl%
\url{https://doi.org/10.3390/ijerph19159295}
\showDOI{\tempurl}


\bibitem[Zheng et~al\mbox{.}(2022)]%
        {zheng2022ai}
\bibfield{author}{\bibinfo{person}{Stephan Zheng}, \bibinfo{person}{Alexander Trott}, \bibinfo{person}{Sunil Srinivasa}, \bibinfo{person}{David~C Parkes}, {and} \bibinfo{person}{Richard Socher}.} \bibinfo{year}{2022}\natexlab{}.
\newblock \showarticletitle{The AI Economist: Taxation policy design via two-level deep multiagent reinforcement learning}.
\newblock \bibinfo{journal}{\emph{Science advances}} \bibinfo{volume}{8}, \bibinfo{number}{18} (\bibinfo{year}{2022}), \bibinfo{pages}{eabk2607}.
\newblock


\end{thebibliography}

\end{document}